\documentclass[a4paper,11pt]{article}
\pdfoutput=1

\usepackage{amsmath,amssymb,bbm,mathtools}
\usepackage{ascmac}
\usepackage{cancel}
\usepackage{xcolor} 
\usepackage[bookmarks=true,bookmarksnumbered=true,bookmarkstype=toc]{hyperref} 
\hypersetup{
pdfauthor={Tetsuji Kimura},
colorlinks={true},
linkcolor={blue},
urlcolor={blue},
filecolor={blue},
citecolor={blue}
}
\definecolor{refkey}{rgb}{0.40, 0.55, 0.55}
\definecolor{labelkey}{rgb}{0.40, 0.55, 0.55}
\usepackage{ulem}
\definecolor{darkcyan}{HTML}{008B8B}

\makeatletter

 \@addtoreset{equation}{section}
\makeatother

\makeatletter
\def\tbcaption{\def\@captype{table}\caption}
\def\figcaption{\def\@captype{figure}\caption}
\makeatother

\newcounter{Enumerate}

\DeclareFontFamily{U}{rsf}{}
\DeclareFontShape{U}{rsf}{m}{n}{
  <5> <6> rsfs5 <7> <8> <9> rsfs7 <10-> rsfs10}{}
\DeclareMathAlphabet\Scr{U}{rsf}{m}{n}

\usepackage[mathscr]{eucal}

\newcommand{\ls}{\ \ \ \ \ }

\newcommand{\bsubeq}{\begin{subequations}}
\newcommand{\esubeq}{\end{subequations}}

\newcommand{\noi}{\noindent}

\newcommand{\eps}{\epsilon}

\newcommand{\nn}{\nonumber}

\newcommand{\I}{{\rm i}}
\newcommand{\T}{{\rm T}}

\renewcommand{\l}{\ell}

\newcommand{\slb}{\scalebox}

\newcommand{\tcc}[1]{\textcolor{red}{$\bullet^{\sf #1}$}}

\def\+{{+\!\!\!+}}

\begin{document}
\allowdisplaybreaks{

\thispagestyle{empty}


\begin{flushright}
TIT/HEP-649 \\
\end{flushright}

\vspace{35mm}

\noi
\slb{2.3}{Supersymmetry Projection Rules}

\vspace{5mm}

\noi
\slb{2.3}{on Exotic Branes}

\vspace{15mm}

\noi
{\renewcommand{\arraystretch}{1.6}
\begin{tabular}{cl}
\multicolumn{2}{l}{\slb{1.2}{Tetsuji {\sc Kimura}} \vphantom{$\Bigg|$}}
\\
& {\renewcommand{\arraystretch}{1.0}
\begin{tabular}{l}
{\sl Research and Education Center for Natural Sciences, Keio University}
\\
{\sl Hiyoshi 4-1-1, Yokohama, Kanagawa 223-8521, JAPAN} 
\end{tabular}
}
\\
& \ls and
\\
& {\renewcommand{\arraystretch}{1.0}
\begin{tabular}{l}
{\sl
Department of Physics,
Tokyo Institute of Technology} \vphantom{$\Big|$}
\\
{\sl Tokyo 152-8551, JAPAN}
\end{tabular}
}
\\
& \ \ \ \slb{0.9}{\tt tetsuji.kimura \_at\_ keio.jp}
\end{tabular}
}


\vspace{35mm}


\begin{abstract}
We study the supersymmetry projection rules on exotic branes in type II string theories and M-theory. They justify the validity of the exotic duality between standard branes and exotic branes of codimension two. By virtue of the supersymmetry projection rules on various branes, we can consider a system 
that involves multiple non-parallel exotic branes
in the nongeometric background. 
\end{abstract}


\newpage
\section{Introduction}
\label{sect:introduction}

Exotic branes \cite{Blau:1997du, Obers:1998fb, Eyras:1999at, LozanoTellechea:2000mc, deBoer:2010ud} should play a central role in the investigation of
the non-perturbative dynamics in string theory and gauge theory.
This is because the exotic branes originate from standard branes such as 
fundamental strings (or F-strings, for short), Neveu-Schwarz fivebranes (NS5-branes) and Dirichlet branes (D-branes) via the string dualities in lower dimensions.
The standard branes have contributed to understanding the non-perturbative dynamics in string theory and gauge theory \cite{Giveon:1998sr}.
However, compared with the standard branes,
the dynamical feature of the exotic branes is still unclear.
The main reason is that the transverse space of an exotic brane has a non-trivial monodromy due to the string dualities \cite{Greene:1989ya, Bergshoeff:2006jj, deBoer:2010ud}.

Consider, for instance, an exotic $5^2_2$-brane.
This object comes from an NS5-brane via the T-duality along two transverse directions of it.
The transverse space has the $SO(2,2;{\mathbb Z}) = SL(2,{\mathbb Z}) \times SL(2,{\mathbb Z})$ monodromy structure which originates from the T-duality \cite{deBoer:2010ud, Kikuchi:2012za, deBoer:2012ma, Kimura:2014wga, Kimura:2014bea, Okada:2014wma}. 
This non-trivial monodromy structure often prevents us from analyzing excitations of the $5^2_2$-brane.
This is completely different from the case of the standard branes\footnote{Strictly speaking, a D7-brane also has
an $SL(2,{\mathbb Z})$ monodromy originating
from the string S-duality \cite{Greene:1989ya}.}.

The exotic branes are also characterized by their masses.
They are different from those of the standard branes.
For instance, the masses of an exotic $b^c_n$-brane and an exotic $b^{(d,c)}_n$-brane are described as
\bsubeq \label{branes-name}
\begin{alignat}{2}
b^c_n : && \ \ \ 
M \ &= \ 
\frac{R_1 \cdots R_b (R_{b+1} \cdots R_{b+c})^2}{g_s^n \l_s^{b+2c+1}}
\, , \\
b^{(d,c)}_n : && \ \ \ 
M \ &= \ 
\frac{R_1 \cdots R_b (R_{b+1} \cdots R_{b+c})^2 (R_{b+c+1} \cdots R_{b+c+d})^3}{g_s^n \l_s^{b+2c+3d+1}}
\, . 
\end{alignat}
\esubeq
Here the labels $b$, $c$ and $n$ indicate
the spatial dimensions, the number of isometry directions in the transverse directions, and the power of the string coupling constant in the tension of the 
exotic brane under consideration, respectively.
The symbol $(d,c)$ also indicates the $c$ and $d$ isometry directions.
$g_s$ and $\l_s$ are the string coupling constant and the string length, respectively.
Each $R_i$ indicates the radius or size in the $i$-th direction.
$R_1 \cdots R_b$ represents the volume of the brane expanded along the spatial $12\cdots b$ directions.
Unlike those of the standard branes,
the mass formulae (\ref{branes-name}) possess multiple powers of radii such as $(R_{b+1})^2$.
Expressions (\ref{branes-name}) are derived from those of the standard branes via the string dualities:
\bsubeq \label{string-dualities}
\begin{alignat}{3}
\text{T$_y$-duality} : \ls & & R_y \ &\to \ \frac{\ell_s^2}{R_y}
\, , \ls & g_s \ &\to \ \frac{\ell_s}{R_y} g_s
\, , \\
\text{S-duality} : \ls & & g_s \ &\to \ \frac{1}{g_s}
\, , \ls & \ell_s \ &\to \ g_s^{1/2} \ell_s
\, .
\end{alignat}
\esubeq
We note that T$_y$ implies the T-duality transformation along the $y$-direction.
For instance, consider the $5^2_2$-brane again.
As mentioned above, this comes from an NS5-brane whose mass is 
\begin{align}
M \ &= \ 
\frac{R_1 R_2 \cdots R_5}{g_s^2 \l_s^6}
\, , \label{mass-NS5}
\end{align}
when the NS5-brane is expanded along the 12345-directions.
Following the nomenclature in \cite{deBoer:2012ma}, 
we refer to this as NS5(12345).
Performing the T-duality along the 89-directions,
we obtain the $5^2_2$(12345,89)-brane whose mass is 
\begin{align}
M \ &= \ 
\frac{R_1 R_2 \cdots R_5 (R_8 R_9)^2}{g_s^2 \l_s^{10}}
\, . \label{mass-522}
\end{align}

Another feature of the exotic branes is that the codimension, i.e., the difference between the bulk spacetime dimensions and the worldvolume dimensions, is less than three.
This implies that the single exotic brane does not have a
well-defined background geometry in the supergravity framework.
For instance, in the case of a standard brane of codimension $k > 3$,
its background geometry is governed by a harmonic function of $r^{-k+2}$,
where $r$ indicates the distance from the core of the brane.
If the codimension is two or one, the harmonic function becomes 
logarithmic or linear, respectively.

Due to the above features, it is often difficult to analyze the global structure of the exotic branes.
However, since the exotic branes are cousins of the standard branes,
their BPS conditions should be characterized in the same way as those of the standard branes.
For instance, a D$p$-brane stretched along the $12\cdots p$ directions preserves supercharges of the form $\eps_L Q_L + \eps_R Q_R$ with 
\begin{align}
\eps_L \ &= \ \Gamma^{012\cdots p} \eps_R
\, , \label{SPR}
\end{align}
where $\eps_L$ and $\eps_R$ are the supersymmetry parameters given by the Majorana-Weyl fermions with 
different chiralities $\Gamma \eps_L = + \eps_L$, $\Gamma \eps_R = - \eps_R$ in type IIA theory, or the same chirality $\Gamma \eps_{L,R} = + \eps_{L,R}$ in type IIB theory. $\Gamma$ is the chirality operator in ten dimensions.
$Q_L$ and $Q_R$ are the corresponding left and right supercharges, and $\Gamma^a$ is the $a$-th Dirac gamma matrix.
The chirality operator $\Gamma$ is described by the Dirac gamma matrices $\Gamma = \Gamma^{012\cdots9}$.
We refer to (\ref{SPR}) as the supersymmetry projection rule.
In this paper, we will explore the supersymmetry projection rules on various exotic branes in type II string theories and M-theory \cite{LozanoTellechea:2000mc, deBoer:2012ma}.


Before moving to the main part of this paper, 
we also mention an interesting relation among defect branes of codimension two \cite{Bergshoeff:2011se}.
There are various defect branes in $D$-dimensional spacetime.
We show them in Table \ref{table:defect-branes}:
\begin{center}
\slb{.9}{\renewcommand{\arraystretch}{1.0}
\begin{tabular}{c|ccccc} \hline
{\sl D}
& $n=0$
& $n=1$
& $n=2$
& $n=3$
& $n=4$
\\ \hline\hline
IIB & & {D7 [$C_8$]} 
& & {${7}_{{3}}$ [$E_8$]} & 
\\
9 & & {D6 [$C_7$]} 
& & {${6}_{{3}}^{{1}}$ [$E_{8,1}$]} & 
\\
8
& & {D5 [$C_6$]} & 
{\renewcommand{\arraystretch}{.8} 
\begin{tabular}{r@{\!\;\ }l} 
{NS5} & [$D_6$] \cr 
KK5 & [$D_{7,1}$] \cr
${5}^{{2}}_{{2}}$ & [$D_{8,2}$] 
\end{tabular}} 
& {${5}_{{3}}^{{2}}$ [$E_{8,2}$]} 
& 
\\
7 & & {D4 [$C_5$]}
& & {${4}_{{3}}^{{3}}$ [$E_{8,3}$]} & 
\\
6 & & {D3 [$C_4$]}
& & {${3}_{{3}}^{{4}}$ [$E_{8,4}$]} & 
\\
5 & & {D2 [$C_3$]}
& & {${2}_{{3}}^{{5}}$ [$E_{8,5}$]} & 
\\
4 & {F1 [$B_2$]}
& {D1 [$C_2$]}
& & {${1}_{{3}}^{{6}}$ [$E_{8,6}$]} 
& {${1}_{{4}}^{{6}}$ [$F_{8,6}$]}
\\
3 & {P} 
& {D0 [$C_1$]}
& & {${0}_{{3}}^{{7}}$ [$E_{8,7}$]} 
& {${0}_{{4}}^{{(1,6)}}$ [$F_{8,7,1}$]}
\\ \hline
\end{tabular}
}
\tbcaption{\small
Defect branes in $D$-dimensional spacetime.
Here the integer $n$ in the first row indicates the power of the string coupling constant in each brane's mass.
D$p$ means the D$p$-brane, while $b^c_n$ and $b^{(d,c)}_n$ represent the exotic branes.
F1 and P denote the F-string and the pp-wave which also behave as the defect branes in four and three dimensions, respectively.
The labels in brackets represent the tensor fields coupling to the corresponding defect branes.
We refer to the branes of $n = 0,1,2$ as fundamental, Dirichlet, and solitonic branes, respectively \cite{Bergshoeff:2011se}.
} 
\label{table:defect-branes}
\end{center}
There exists 
an $SL(2,{\mathbb Z})$ duality group
under which the standard branes of $n = 0,1$ are mapped to the exotic branes of $n = 4,3$ and vice versa, and the solitonic branes of $n = 2$ are mapped to other solitonic branes.
This duality group is a subgroup of the U-duality group in each dimension.
This is referred to as the {\it exotic duality} \cite{Bergshoeff:2011se, Sakatani:2014hba}.
Even though the U-duality group in a certain spacetime dimension
is different from that of a different dimension,
any exotic duality is described by $SL(2,{\mathbb Z})$.
This duality is illustrated in Figures \ref{fig:exotic-dual-pbranes} and \ref{fig:exotic-dual-5branes}:
\begin{center}
\slb{.85}{\includegraphics[bb=0 0 368 100]{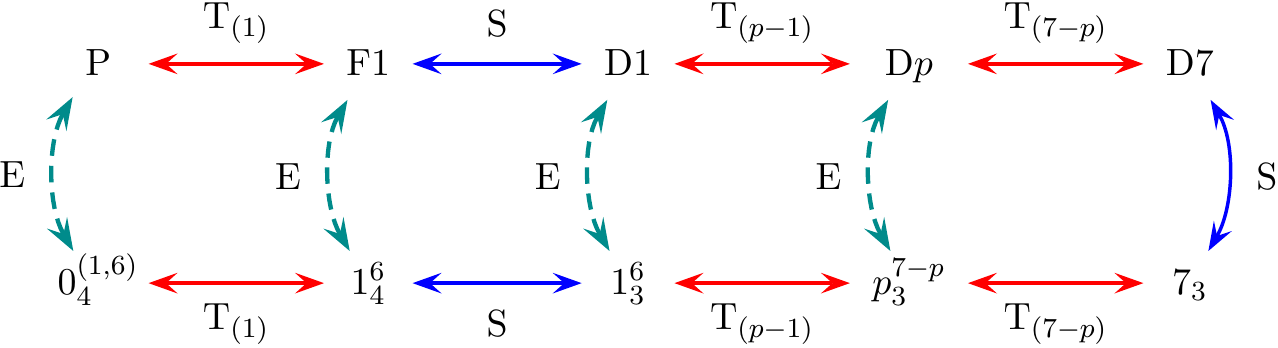}} 
\figcaption{\small 
Exotic duality, S-duality and T-duality along $k$ directions (labeled as E, S and T$_{(k)}$, respectively) among the standard branes and the exotic branes \cite{Bergshoeff:2011se, Sakatani:2014hba}.
}
\label{fig:exotic-dual-pbranes}
\end{center}
%
\begin{center}
\slb{.85}{\includegraphics[bb=0 0 189 95]{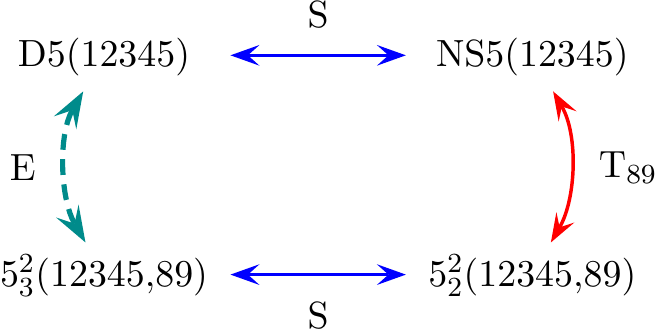}}
\figcaption{\small
Exotic duality between D5-brane and $5^2_3$-brane.
The terminology T$_{89}$ implies that the T-duality is performed along the 89-directions.
It is remarked that this T$_{89}$-duality between NS5-brane and $5^2_2$-brane is also interpreted as an exotic duality \cite{Bergshoeff:2011se, Sakatani:2014hba}.
}
\label{fig:exotic-dual-5branes}
\end{center}

The exotic duality of a {\it single} brane was suggested in \cite{Bergshoeff:2011se} from the viewpoint of the string duality groups and their representations.
This was also analyzed in \cite{Kleinschmidt:2011vu}
by virtue of the $E_{11}$ supergravity technique.
Furthermore, in the framework of other extended supergravity such as $\beta$-supergravity \cite{Andriot:2013xca, Andriot:2014uda} and its extended version \cite{Blair:2014zba, Sakatani:2014hba}\footnote{The extended version of $\beta$-supergravity \cite{Blair:2014zba, Sakatani:2014hba} involves the Ramond-Ramond potentials $C_p$ and their string dualized objects $\gamma^p$. This formulation may be referred to as ``$\gamma$-supergravity''.},
the exotic duality was further investigated.
In this work, we would like to confirm the validity of the exotic duality from the viewpoint of the supersymmetry projection rules such as (\ref{SPR}) 
discussed in \cite{LozanoTellechea:2000mc, deBoer:2012ma}, 
and apply it to new brane configurations that involve {\it multiple} non-parallel (exotic) branes.

\vspace{1mm}

The structure of this paper is as follows.
In section \ref{sect:SPR},
we first list the supersymmetry projection rules on the standard branes.
Extracting the string dualities acting on the supersymmetry parameters, 
we explicitly write down the supersymmetry projection rules on the exotic branes.
We find that their expressions justify the exotic duality from the supersymmetry viewpoint.
In order to check the consistency,
we apply the supersymmetry projection rules on the exotic branes to certain brane configurations that contain exotic branes.
In section \ref{sect:ED-BEB},
we consider the exotic duality applied to multiple non-parallel branes.
By virtue of the supersymmetry projection rules discussed in section \ref{sect:SPR}, we find various interesting configurations.
Section \ref{sect:discussions} is devoted to the conclusion and discussions.
In appendix \ref{app:data} the detailed computations to derive the supersymmetry projection rules on various exotic branes are explicitly described.

\section{Supersymmetry projection rules}
\label{sect:SPR}

In this section, 
we first exhibit the supersymmetry projection rules such as (\ref{SPR}) on the standard branes in type II string theories and M-theory.
Following the rules, we introduce the string dualities acting on the supersymmetry parameters.
Using the string dualities, we write down the rules on various exotic branes.
To avoid complications, we do not write down the concrete derivation of each exotic brane in this section.
It is summarized in appendix \ref{app:data}.
Next, we apply the supersymmetry projection rules to certain brane configurations derived from an F-string ending on a D3-brane.
Analogous to the string dualities on the mass formulae of branes (\ref{string-dualities}),
we do not seriously consider their global structures.

\subsection{Rules on standard branes}
\label{sect:SPR-standard}

First of all, we gather the supersymmetry projection rules on standard branes.
These are very common and can be seen in the literature (see, for instance, \cite{Gauntlett:1997cv, Giveon:1998sr}):
\begin{itemize}
\item Standard branes in IIA theory:
\bsubeq \label{SPR-IIA-standard}
\begin{alignat}{2}
&&
\Gamma \eps \ &= \ \sigma_3 \, \eps
\, , \ \ \ \text{or equivalently} \ \ \
\Gamma \eps_L \ = \ + \eps_L
\, , \ \ \ 
\Gamma \eps_R \ = \ - \eps_R
\, , \\
\text{P(1)}: && \ \ \ 
\pm \eps \ &= \ \Gamma^{01} \eps 
\, , \\
\text{F1(1)}: && \ \ \ 
\pm \eps \ &= \ \Gamma^{01} \Gamma \eps
\, , \ls
\text{NS5(12345)}: \ \ \ 
\pm \eps \ = \ \Gamma^{012345} \eps
\, , \\
\text{D$p$($12\cdots p$)}: && \ \ \ 
\pm \eps \ &= \ 
\Gamma^{012 \cdots p} A_p (\sigma_1) \eps
\, , \ls
A_p \ \equiv \ \left\{
\begin{array}{rl}
1 : & p = 2,6,
\\
\Gamma : & p = 0,4,8.
\end{array}
\right.
\end{alignat}
\esubeq

\item Standard branes in IIB theory:
\bsubeq \label{SPR-IIB-standard}
\begin{alignat}{2}
&&
\Gamma \eps \ &= \ \eps
\, , \ \ \ \text{or equivalently} \ \ \
\Gamma \eps_L \ = \ + \eps_L
\, , \ \ \ 
\Gamma \eps_R \ = \ + \eps_R
\, , \\
\text{P(1)}: && \ \ \ 
\pm \eps \ &= \ \Gamma^{01} \eps 
\, , \\
\text{F1(1)}: && \ \ \ 
\pm \eps \ &= \ \Gamma^{01} (\sigma_3) \eps
\, , \ls
\text{NS5(12345)}: \ \ \ 
\pm \eps \ = \ \Gamma^{012345} (\sigma_3) \eps
\, , \\
\text{D$p$($12\cdots p$)}: && \ \ \ 
\pm \eps \ &= \ \Gamma^{012 \cdots p} B_p \eps
\, , \ls
B_p \ \equiv \ 
\left\{
\begin{array}{rl}
\sigma_1 : & p = 1,5,9,
\\
\I \sigma_2 : & p = 3,7.
\end{array}
\right.
\end{alignat}
\esubeq

\item Standard branes in M-theory:
\bsubeq \label{SPR-M-standard}
\begin{alignat}{2}
\text{M2(12)}: && \ \ \ 
\pm \eta \ &= \ \Gamma^{012} \eta
\, , \\
\text{M5(12345)}: && \ \ \ 
\pm \eta \ &= \ \Gamma^{012345} \eta
\, . 
\end{alignat}
\esubeq

\end{itemize}

In order to make the discussion clear, we introduce
a ``doublet'' of two supersymmetry parameters $\eps_L$ and $\eps_R$
in such a way that $\eps \equiv (\eps_L, \eps_R)^{\T}$.
This is a Majorana spinor in type IIA theory,
while it can be interpreted as an $SL(2,{\mathbb Z})$ doublet in type IIB theory.
Due to this description, we also introduce
the Pauli matrices $\sigma_i$ acting on the doublet $\eps$.
We also note that $\Gamma$ is the chirality operator defined as $\Gamma = \Gamma^{0123456789}$ in terms of the Dirac gamma matrices $\Gamma^a$.
They are subject to the Clifford algebra $\{ \Gamma^a, \Gamma^b \} = 2 \eta^{ab}$.
In this work we use the mostly plus signature $\eta_{ab} = {\rm diag}.(-++\cdots+)$.
In the case of M-theory, the supersymmetry parameter $\eta$ is a Majorana spinor. Dirac gamma matrices in eleven dimensions are the same as those in ten dimensions, while the chirality operator $\Gamma$ is uplifted to the eleventh gamma matrix $\Gamma^{\natural}$ in eleven-dimensional theory.

\subsection{Dualities on supersymmetry parameters}
\label{sect:duality2SUSY}

Since various standard branes are related to each other via the string dualities,
there should exist duality transformation rules on the supersymmetry parameters $\eps_L$ and $\eps_R$.
Here, without the derivation, we exhibit the T-duality and S-duality 
\cite{Schwarz:1983qr}
(for instance, see \cite{Imamura}):
\bsubeq \label{TS-rules-SUSY}
\begin{alignat}{2}
\text{T$_y$-duality}: && \ \ \ 
\eps_L \ &\to \ \eps_L
\, , \ls
\eps_R \ \to \ \Gamma^y \Gamma \eps_R
\, , \label{T-dual-rule} \\
\text{S-duality}: && \ \ \ 
\eps \ &\to \ S \eps
\, , \ls
S \ = \ 
\frac{1}{\sqrt{2}} \big( {\mathbbm 1}_2 - \I \sigma_2 \big)
\, . \label{S-dual-rule}
\end{alignat}
\esubeq
We have a couple of comments on the above rules.
In the case of the T$_y$-duality, the operator $\Gamma^y \Gamma$ generates the parity transformation along the $y$-direction such as 
$(\Gamma^y \Gamma)^{-1} \Gamma^y (\Gamma^y \Gamma) = - \Gamma^y$,
while this behaves as an identity operator acting on the other $\Gamma^i$ ($i \neq y$) as follows:
$(\Gamma^y \Gamma)^{-1} \Gamma^i (\Gamma^y \Gamma) = \Gamma^i$.
In the S-duality case, this rule does not change the supersymmetry projection rule on the D3-brane.
This is expressed by $S^{-1} (\I \sigma_2) S = \I \sigma_2$.
On the other hand, the operator $S$ transforms $\sigma_1$ and $\sigma_3$ in such a way that $S^{-1} \sigma_1 S = \sigma_3$ and $S^{-1} \sigma_3 S = - \sigma_1$.
Under this transformation, the F-string and the NS5-brane are mapped to the D-string and the D5-brane, and vice versa.
Geometrically, the S-duality transformation means a rotation along the second axis of the three-dimensional $SL(2,{\mathbb Z})$ space.

\subsection{Rules on exotic branes}
\label{sect:result}

Applying the string dualities to the supersymmetry projection rules on the standard branes, 
we can derive those of various exotic branes in a straightforward way.
As mentioned before, the explicit computations are listed in appendix \ref{app:data}.
Here we summarize the supersymmetry projection rules on 
the solitonic branes and 
defect branes discussed in \cite{LozanoTellechea:2000mc, deBoer:2012ma}, and the rules on the domain walls which are new.
First we exhibit the rules on the solitonic branes in type IIA and IIB theories, respectively.
\begin{itemize}
\item Solitonic five-branes in IIA theory:
\bsubeq \label{SPR-IIA-5}
\begin{alignat}{2}
\text{NS5}(12345) : && \ \ \ 
\pm \eps \ &= \ 
\Gamma^{012345} \eps
\, , \\
\text{KK5}(12345,9) : && \ \ \ 
\pm \eps \ &= \ 
\Gamma^{012345} \Gamma \eps
\, , \\
5^2_2 (12345,89) : && \ \ \ 
\pm \eps \ &= \ 
\Gamma^{012345} \eps
\, , \\
5^3_2 (12345,789) : && \ \ \ 
\pm \eps \ &= \ 
\Gamma^{012345} \Gamma \eps
\, , \\
5^4_2 (12345,6789) : && \ \ \ 
\pm \eps \ &= \ 
\Gamma^{012345} \eps
\, .
\end{alignat}
\esubeq

\item Solitonic five-branes in IIB theory:
\bsubeq \label{SPR-IIB-5}
\begin{alignat}{2}
\text{NS5}(12345) : && \ \ \ 
\pm \eps \ &= \ 
\Gamma^{012345} (\sigma_3) \eps
\, , \\
\text{KK5}(12345,9) : && \ \ \ 
\pm \eps \ &= \ 
\Gamma^{012345} \eps
\, , \\
5^2_2 (12345,89) : && \ \ \ 
\pm \eps \ &= \ 
\Gamma^{012345} (\sigma_3) \eps
\, , \\
5^3_2 (12345,789) : && \ \ \ 
\pm \eps \ &= \ 
\Gamma^{012345} \eps
\, , \\
5^4_2 (12345,6789) : && \ \ \ 
\pm \eps \ &= \ 
\Gamma^{012345} (\sigma_3) \eps
\, .
\end{alignat}
\esubeq
We find that the transverse directions with isometry (i.e., the 6789-directions) do not contribute to the supersymmetry projection rules.
On the other hand, analogous to the NS5-brane, the hyperplanes in which the five-branes are stretched provide the projections.
The above expressions guarantee that the {\it defect} $(p,q)$ five-brane, a bound state of a defect NS5(12345) and an exotic $5^2_2(12345,89)$, is a 1/2-BPS object in string theory \cite{Kimura:2014wga, Kimura:2014bea, Okada:2014wma},
while a bound state of a KK5(12345,9) and a $5^2_2(12345,89)$ breaks supersymmetry.
\end{itemize}


Next, we gather the supersymmetry projection rules on the defect branes \cite{deBoer:2010ud, Bergshoeff:2011se, deBoer:2012ma} in Table \ref{table:defect-branes}.
\begin{itemize}
\item Defect branes in IIA theory:
\bsubeq \label{SPR-IIA-defect}
\begin{alignat}{2}
6^1_3 (123456,7) : && \ \ \ 
\pm \eps \ &= \ 
\Gamma^{0123456} (\sigma_1) \eps
\, , \\
4^3_3 (1234,567) : && \ \ \ 
\pm \eps \ &= \ 
\Gamma^{01234} \Gamma (\sigma_1) \eps
\, , \\
2^5_3 (12,34567) : && \ \ \ 
\pm \eps \ &= \ 
\Gamma^{012} (\sigma_1) \eps
\, , \\
0^7_3 (,1234567) : && \ \ \ 
\pm \eps \ &= \ 
\Gamma^{0} \Gamma (\sigma_1) \eps
\, , \\
1^6_4 (1,234567) : && \ \ \ 
\pm \eps \ &= \ 
\Gamma^{01} \Gamma \eps
\, , \\
0^{(1,6)}_4 (,234567,1) : && \ \ \ 
\pm \eps \ &= \ 
\Gamma^{01} \eps
\, .
\end{alignat}
\esubeq

\item Defect branes in IIB theory:
\bsubeq \label{SPR-IIB-defect}
\begin{alignat}{2}
7_3 (1234567) : && \ \ \ 
\pm \eps \ &= \ 
\Gamma^{01234567} (\I \sigma_2) \eps
\, , \\
5^2_3 (12345,67) : && \ \ \ 
\pm \eps \ &= \ 
\Gamma^{012345} (\sigma_1) \eps
\, , \\
3^4_3 (123,4567) : && \ \ \ 
\pm \eps \ &= \ 
\Gamma^{0123} (\I \sigma_2) \eps
\, , \\
1^6_3 (1,234567) : && \ \ \ 
\pm \eps \ &= \ 
\Gamma^{01} (\sigma_1) \eps
\, , \\
1^6_4 (1,234567) : && \ \ \ 
\pm \eps \ &= \ 
\Gamma^{01} (\sigma_3) \eps
\, , \\
0^{(1,6)}_4 (,234567,1) : && \ \ \ 
\pm \eps \ &= \ 
\Gamma^{01} \eps
\, .
\end{alignat}
\esubeq
We have summarized the supersymmetry projection rules on all the defect branes in Table \ref{table:defect-branes}.
We have comments on the defect branes.
The supersymmetry projection rule on each defect $p^{7-p}_3$-brane coincides with that of the D$p$-brane. These two branes are exotic dual.
The rules on the $1^6_4$-brane and the $0^{(1,6)}_4$-brane in type IIA/IIB theories are also exactly the same as those of the F-string and the pp-wave, respectively. The former exotic branes are exotic dual of the latter standard branes \cite{Bergshoeff:2011se, Sakatani:2014hba}.
\end{itemize}


Applying the additional string dualities to the defect branes in Table \ref{table:defect-branes}, 
we obtain the domain walls represented as $b^{(1,c)}_3$-branes and $b^{(d,3)}_4$-branes \cite{Bergshoeff:2012pm}.
\begin{itemize}
\item Domain walls in IIA theory:
\bsubeq \label{SPR-IIA-DW}
\begin{alignat}{2}
7^{(1,0)}_3 (1234567,,9) : && \ \ \ 
\pm \eps \ &= \ 
\Gamma^{012345679} \Gamma (\sigma_1) \eps
\, , \\
5^{(1,2)}_3 (12345,67,9) : && \ \ \ 
\pm \eps \ &= \ 
\Gamma^{0123459} (\sigma_1) \eps
\, , \\
3^{(1,4)}_3 (123,4567,9) : && \ \ \ 
\pm \eps \ &= \ 
\Gamma^{01239} \Gamma (\sigma_1) \eps
\, , \\
1^{(1,6)}_3 (1,234567,9) : && \ \ \ 
\pm \eps \ &= \ 
\Gamma^{019} (\sigma_1) \eps
\, , \\
5^3_4 (12345,789) : && \ \ \ 
\pm \eps \ &= \ 
\Gamma^{012345} \eps
\, , \\
4^{(1,3)}_4 (1234,789,5) : && \ \ \ 
\pm \eps \ &= \ 
\Gamma^{012345} \Gamma \eps
\, , \\
3^{(2,3)}_4 (123,789,45) : && \ \ \ 
\pm \eps \ &= \ 
\Gamma^{012345} \eps
\, , \\
2^{(3,3)}_4 (12,789,345) : && \ \ \ 
\pm \eps \ &= \ 
\Gamma^{012345} \Gamma \eps
\, , \\
1^{(4,3)}_4 (1,789,2345) : && \ \ \ 
\pm \eps \ &= \ 
\Gamma^{012345} \eps
\, , \\
0^{(5,3)}_4 (,789,12345) : && \ \ \ 
\pm \eps \ &= \ 
\Gamma^{012345} \Gamma \eps
\, .
\end{alignat}
\esubeq

\item Domain walls in IIB theory:
\bsubeq \label{SPR-IIB-DW}
\begin{alignat}{2}
6^{(1,1)}_3 (123456,7,9) : && \ \ \ 
\pm \eps \ &= \ 
\Gamma^{01234569} (\I \sigma_2) \eps
\, , \\
4^{(1,3)}_3 (1234,567,9) : && \ \ \ 
\pm \eps \ &= \ 
\Gamma^{012349} (\sigma_1) \eps
\, , \\
2^{(1,5)}_3 (12,34567,9) : && \ \ \ 
\pm \eps \ &= \ 
\Gamma^{0129} (\I \sigma_2) \eps
\, , \\
0^{(1,7)}_3 (,1234567,9) : && \ \ \ 
\pm \eps \ &= \ 
\Gamma^{09} (\sigma_1) \eps
\, , \\
5^3_4 (12345,789) : && \ \ \ 
\pm \eps \ &= \ 
\Gamma^{012345} \eps
\, , \\
4^{(1,3)}_4 (1234,789,5) : && \ \ \ 
\pm \eps \ &= \ 
\Gamma^{012345} (\sigma_3) \eps
\, , \\
3^{(2,3)}_4 (123,789,45) : && \ \ \ 
\pm \eps \ &= \ 
\Gamma^{012345} \eps
\, , \\
2^{(3,3)}_4 (12,789,345) : && \ \ \ 
\pm \eps \ &= \ 
\Gamma^{012345} (\sigma_3) \eps
\, , \\
1^{(4,3)}_4 (1,789,2345) : && \ \ \ 
\pm \eps \ &= \ 
\Gamma^{012345} \eps
\, , \\
0^{(5,3)}_4 (,789,12345) : && \ \ \ 
\pm \eps \ &= \ 
\Gamma^{012345} (\sigma_3) \eps
\, .
\end{alignat}
\esubeq
We have comments on the rules on the domain walls.
Analogous to D-branes,
there exist BPS $(2b+1)^{(1,c)}_3$-branes in type IIA theory, 
and BPS $(2b)^{(1,c)}_3$-branes in type IIB theory.
On the other hand, 
there exist BPS $b^{(d,3)}_4$-branes with $b+d = 5$ in both IIA and IIB theories.
It is noticed that the above is not a complete list of domain walls.
There exist other kinds of domain walls in type II string theories.
For instance, we find a type IIB $5^4_5(12345,6789)$-brane 
originating from a type IIB NS5(12345)-brane via the string ST$_{6789}$-dualities.
More complicatedly, a type IIB $3^{(1,2,3)}_4(123,678,45,9)$-brane can be derived from the type IIB NS5(12345)-brane under the string T$_{459}$ST$_{678}$-dualities.
In principle, we can obtain the supersymmetry projection rules on the $5^4_5$-brane and the $3^{(1,2,3)}_4$-brane under the duality rules (\ref{TS-rules-SUSY}), in the same way as the derivation of their masses by using (\ref{string-dualities}).
\end{itemize}

As listed above, the supersymmetry projection rules on the exotic branes are quite similar to those of the standard branes.
In particular, we emphasize that the rules on the defect branes (\ref{SPR-IIA-defect}) and (\ref{SPR-IIB-defect}) coincide with those of the standard branes (\ref{SPR-IIA-standard}) and (\ref{SPR-IIB-standard}), respectively.
Hence they justify the validity of the exotic duality illustrated in Figures \ref{fig:exotic-dual-pbranes} and \ref{fig:exotic-dual-5branes} from the supersymmetry viewpoint.

\subsection{Exotic branes in M-theory}
\label{sect:exoticM}

Once we have understood the supersymmetry projection rules in type IIA theory,
we can uplift them to those in M-theory.
This procedure is simple because the type IIA supersymmetry parameter
$\eps$ becomes a Majorana spinor $\eta$ in M-theory.
Uplifting type IIA theory to M-theory,
we can also interpret the chirality operator $\Gamma$ as the eleventh Dirac gamma matrix $\Gamma^{\natural}$.
Furthermore, we have also identified the relation between
the string coupling $g_s$, the string length $\ell_s$, the radius of the M-theory circle $R_{\natural}$, and the eleven-dimensional Planck length $\ell_P$ in the literature:
\bsubeq \label{mass-IIA2M}
\begin{alignat}{2}
\eps \ &= \ \eta
\, , &\ls
\Gamma \ &= \ \Gamma^{\natural} 
\, , \\
g_s \ell_s \ &= \ R_{\natural} 
\, , &\ls
g_s^{1/3} \ell_s \ &= \ \ell_P 
\, . 
\end{alignat}
\esubeq
Applying the uplift to various exotic branes in type IIA theory,
we obtain the exotic branes and their supersymmetry projection rules:
\bsubeq \label{SPR-M-defect}
\begin{alignat}{2}
\text{KK6(12345$\natural$,9)}: && \ \ \ 
\pm \eta \ &= \ 
\Gamma^{012345\natural} \eta
\, , \\
5^3(12345,89\natural) : && \ \ \ 
\pm \eta \ &= \ 
\Gamma^{012345} \eta
\, , \\
5^{(1,3)}(12345,789,\natural) : && \ \ \ 
\pm \eta \ &= \ 
\Gamma^{012345\natural} \eta
\, , \\
2^6(1\natural,234567) : && \ \ \ 
\pm \eta \ &= \ 
\Gamma^{01\natural} \eta
\, , \\
0^{(1,7)}(,234567\natural,1) : && \ \ \ 
\pm \eta \ &= \ 
\Gamma^{01} \eta
\, .
\end{alignat}
\esubeq
We again summarize
the detailed computations in appendix \ref{app:IIA2M}.
Here, for simplicity, we skip consideration of the uplift of the domain walls in (\ref{SPR-IIA-DW}).

\subsection{Brane configurations as a consistency check}
\label{sect:BC}

In order to check the consistency of the supersymmetry projection rules on the exotic branes,
we apply them to a certain brane configuration.
In this paper we focus on the system in which a brane is ending on another brane.
A typical example is the system of an F-string ending on a D3-brane, shown in Figure \ref{fig:D3-F1}:
\begin{center}
\begin{tabular}{cc}
\slb{.85}{\includegraphics[bb=0 0 84 74]{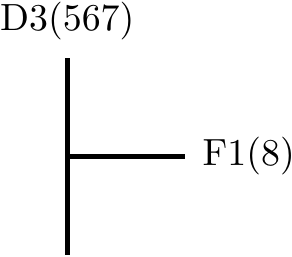}}
&
\raisebox{10mm}{\slb{.9}{\renewcommand{\arraystretch}{1.3}
\begin{tabular}{c||c|cccc|ccc|c|c} \hline
IIB & 0 & 1 & 2 & 3 & 4 & 5 & 6 & 7 & 8 & 9
\\ \hline\hline
D3 & $-$ & & & & & $-$ & $-$ & $-$ & &
\\
F1 & $-$ & & & & &&&& $-$ &
\\ \hline
\end{tabular}
}}
\end{tabular}
\figcaption{\small 
F-string ending on D3-brane. 
The F-string is stretched along the 8-th direction, 
while the D3-brane is expanded in the 567-plane.
}
\label{fig:D3-F1}
\end{center}
It is easy to confirm that the supersymmetry projection rules on the D3-brane and the F-string (\ref{SPR-IIB-standard}) preserve a quarter of the supersymmetry of the system in Figure \ref{fig:D3-F1}.
Next, we apply the string dualities to the system in Figure \ref{fig:D3-F1} and obtain various configurations.
Performing the S-duality and the T-dualities along the 1234-directions, 
followed by the S-duality again,
we obtain the configuration in which the $7_3$-brane is involved (see Figure \ref{fig:73-NS5}):
\begin{center}
\begin{tabular}{cc}
\slb{.85}{\includegraphics[bb=0 0 131 74]{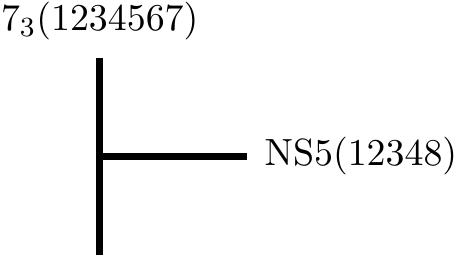}}
&
\raisebox{10mm}{\slb{.9}{\renewcommand{\arraystretch}{1.3}
\begin{tabular}{c||c|cccc|ccc|c|c} \hline
IIB & 0 & 1 & 2 & 3 & 4 & 5 & 6 & 7 & 8 & 9
\\ \hline\hline
$7_3$ & $-$ & $-$ & $-$ & $-$ & $-$ & $-$ & $-$ & $-$ & &
\\
NS5 & $-$ & $-$ & $-$ & $-$ & $-$ &&&& $-$ &
\\ \hline
\end{tabular}
}}
\end{tabular}
\figcaption{\small 
NS5-brane ending on $7_3$-brane. This is ST$_{1234}$S-dual of the configuration in Figure \ref{fig:D3-F1}.
}
\label{fig:73-NS5}
\end{center}
We note that the $7_3$-brane is also called the NS7-brane or the (0,1) sevenbrane.
Since the configuration in Figure \ref{fig:73-NS5} is derived from that in Figure \ref{fig:D3-F1} via the string ST$_{1234}$S-dualities, this should also preserve the 1/4-BPS condition.
This can be confirmed by using the supersymmetry projection rules on the IIB NS5-brane in (\ref{SPR-IIB-standard}) and the $7_3$-brane in (\ref{SPR-IIB-defect}).
The former is given as $\eps = \Gamma^{012348} \eps$ and the latter is $\eps = \Gamma^{01234567} \eps$.
These two conditions provide the equation $\eps = \Gamma^{5678} \eps$.
Since $\Gamma^{5678}$ is traceless and its square becomes the identity,
we can choose the $+1$ eigenvalue of the supersymmetry parameter $\eps$.
Thus it turns out that the configuration preserves a quarter of the supersymmetry.

We can consider a more complicated brane configuration.
Applying the ST$_{4679}$-dualities to the system in Figure \ref{fig:73-NS5}, 
we obtain the brane configuration in which 
the domain wall $5^3_4$-brane is ending on the domain wall $4^{(1,3)}_4$-brane
(see Figure \ref{fig:4,13,4-534}):
\begin{center}
\begin{tabular}{cc}
\slb{.85}{\includegraphics[bb=0 0 153 78]{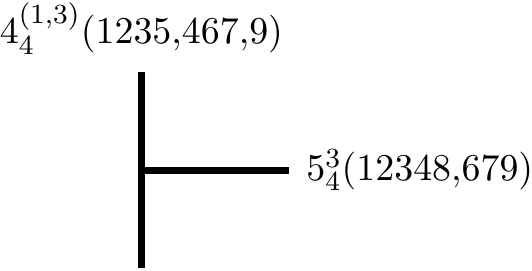}}
&
\raisebox{10mm}{\slb{.9}{\renewcommand{\arraystretch}{1.3}
\begin{tabular}{c||c|cccc|ccc|c|c} \hline
IIB & 0 & 1 & 2 & 3 & 4 & 5 & 6 & 7 & 8 & 9
\\ \hline\hline
$4^{(1,3)}_4$ & $-$ & $-$ & $-$ & $-$ & \tcc{2} & $-$ & \tcc{2} & \tcc{2} & & \tcc{3}
\\
$5^3_4$ & $-$ & $-$ & $-$ & $-$ & $-$ & & \tcc{2} & \tcc{2} & $-$ & \tcc{2}
\\ \hline
\end{tabular}
}}
\end{tabular}
\figcaption{\small
$5^3_4$-brane ending on $4^{(1,3)}_4$-brane. This is ST$_{4679}$-dual of the configuration in Figure \ref{fig:73-NS5}.
The symbols \tcc{2} and \tcc{3} in the table imply that 
the mass of the brane depends on the corresponding direction with the describing power (see the general mass formulae (\ref{branes-name})).
}
\label{fig:4,13,4-534}
\end{center}
We consider the supersymmetry projection rules on the type IIB $4^{(1,3)}_4$-brane and the $5^3_4$-brane in (\ref{SPR-IIB-DW}).
The former is 
$\eps = \Gamma^{012359} (\sigma_3) \eps$, 
while the latter is $\eps = \Gamma^{012348} \eps$. 
Splitting them into equations for $\eps_L$ and $\eps_R$, 
we obtain $\eps_L = - \Gamma^{4589} \eps_L$ and 
$\eps_R = + \Gamma^{4589} \eps_R$.
Since $\Gamma^{4589}$ is traceless and its square is the identity,
we can choose the eigenvalues $-1$ and $+1$ on the parameters $\eps_L$ and $\eps_R$, respectively. 
Hence we can again find that the configuration preserves a quarter of the supersymmetry in type IIB theory.

\section{Exotic mappings of multiple branes}
\label{sect:ED-BEB}

In the previous section, we established the supersymmetry projection rules on the exotic branes as well as those of the standard branes.
As mentioned before, 
the exotic duality was first discussed in \cite{Bergshoeff:2011se} and further developed in \cite{Sakatani:2014hba}.
This is the duality between a single standard brane and a single exotic brane, as illustrated in Figures \ref{fig:exotic-dual-pbranes} and \ref{fig:exotic-dual-5branes}.
In this section, we explore 
a mapping
of {\it multiple} non-parallel (exotic) branes.
This is a procedure in which a certain brane configuration in the conventional framework is mapped to a new brane configuration 
that is realized in the nongeometric framework. The new configuration cannot be described in ordinary supergravity theories. It should be given in terms of the $\beta$-supergravity or its extended version which govern the string theory of nongeometric backgrounds \cite{Andriot:2013xca, Andriot:2014uda, Blair:2014zba, Sakatani:2014hba}.
From now on, we refer to this mapping as exotic mapping.

First we apply the exotic 
mapping of
a D$p$-brane ending on a D$(p+2)$-brane in Figure \ref{fig:exotic1}:
\begin{center}
\slb{.85}{\includegraphics[bb=0 0 279 76]{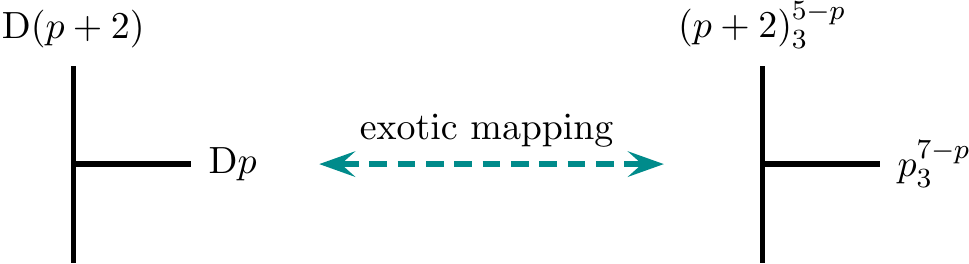}}
\figcaption{\small Exotic mapping from D$(p+2)$--D$p$ system to $(p+2)^{5-p}_3$--$p^{7-p}_3$ system.}
\label{fig:exotic1}
\end{center}
Under the exotic duality, the D$p$-brane and the D$(p+2)$-brane are mapped to 
the $p^{7-p}_3$-brane and the $(p+2)^{5-p}_3$-brane, respectively.
Here we should notice that the exotic duality from D$p$ to $p^{7-p}_3$ is given as the T$_{(7-p)}$ST$_{(7-p)}$-dualities, 
while the exotic duality from D$(p+2)$ to $(p+2)^{5-p}_3$ is 
the T$_{(5-p)}$ST$_{(5-p)}$-dualities\footnote{As introduced in section \ref{sect:introduction}, the terminology $T_{(k)}$ indicates the T-duality along $k$ directions.}, as illustrated in Figure \ref{fig:exotic-dual-pbranes}.
We should notice that we cannot map the left configuration in Figure \ref{fig:exotic1} to the right one via the string dualities. The configuration in the right figure is realized only in the nongeometric background.
However, since we have already understood that the supersymmetry projection rules on the D$p$-brane and the D$(p+2)$-brane are exactly same as those of the $p^{7-p}_3$-brane and the $(p+2)^{5-p}_3$-brane, 
we would be able to 
consider the existence of the right configuration in Figure \ref{fig:exotic1} in the nongeometric framework.

Similarly, we can also consider the exotic 
mapping
of the configuration in which a solitonic brane is ending on a D$p$-brane as in Figure \ref{fig:exotic2}:
\begin{center}
\slb{.85}{\includegraphics[bb=0 0 265 76]{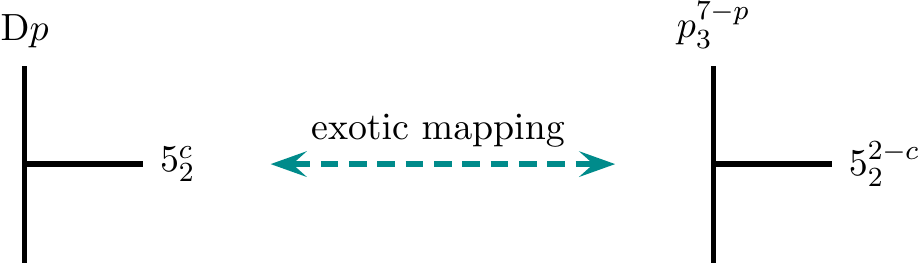}}
\figcaption{\small 
Exotic mapping from D$p$--$5^c_2$ system to $p^{7-p}_3$--$5^{2-c}_2$ system, where $c$ is restricted to $c = 0,1,2$.
}
\label{fig:exotic2}
\end{center}
Here the integer $c$ is restricted to $c = 0,1,2$ in order to avoid the emergence of a domain wall.
This map would also be
applicable because the supersymmetry projection rule on the $5^c_2$-brane coincides with that of the $5^{2-c}_2$-brane.
Furthermore, we can consider the exotic 
mapping
of the D3--F1 system in Figure \ref{fig:exotic3}:
\begin{center}
\slb{.85}{\includegraphics[bb=0 0 254 75]{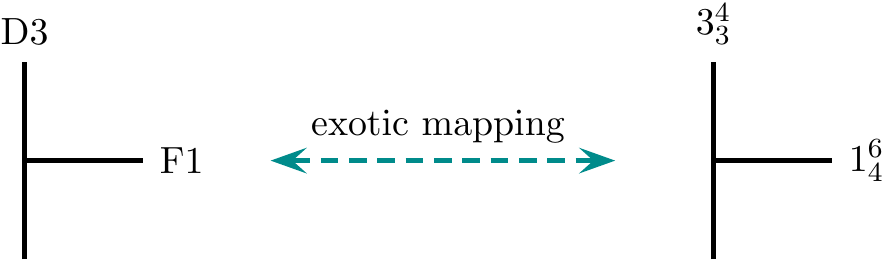}}
\figcaption{\small Exotic mapping from D3--F1 system to $3^4_3$--$1^6_4$ system.}
\label{fig:exotic3}
\end{center}
Again the supersymmetry projection rule on the $1^6_4$-brane is equal to that of the F-string.
Once we recognize the validity of the exotic 
mapping
in Figure \ref{fig:exotic3}, we can immediately apply the T-dualities to this system and obtain the configurations in Figure \ref{fig:exotic4}:
\begin{center}
%
\slb{.85}{\includegraphics[bb=0 0 254 76]{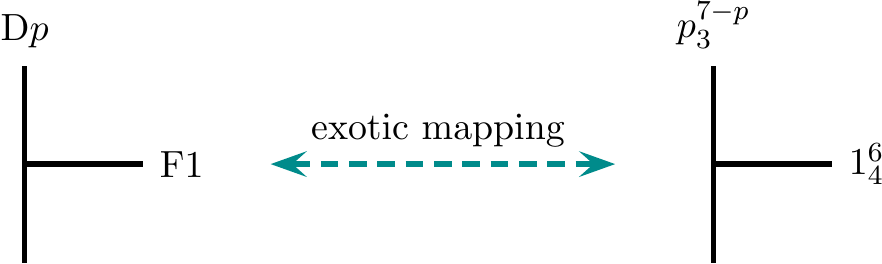}}
\figcaption{\small Exotic mapping from D$p$--F1 system to $p^{7-p}_3$--$1^6_4$ system.}
\label{fig:exotic4}
\end{center}
Since both the F-string and the $1^6_4$-brane exist in type IIA theory as well as in type IIB theory,
the exotic 
mapping
of any integer $p$ in Figure \ref{fig:exotic4} is applicable.

We should be able to consider a certain property from the two configurations in Figure \ref{fig:exotic4}.
In the left picture, we can read off the excitations of the D$p$-brane in terms of the mode excitations of the (open) F-string ending on the D$p$-brane in the small string coupling regime $g_s \to 0$.
By the same analogy, 
we should be able to read off the excitations of the defect $p^{7-p}_3$-brane in terms of the ``mode excitations'' of the (open) defect $1^6_4$-brane in the strong coupling limit $g_s \to \infty$.
Unfortunately, however, we have not understood any mode excitations of the $1^6_4$-brane.
Thus, it seems quite difficult to evaluate the excitations of the defect $p^{7-p}_3$-branes with our current understanding.

We have studied the validity of the exotic mapping applied to the configurations of 
multiple non-parallel (exotic) branes. 
In order to prove this validity completely, 
we have to take care of the non-trivial monodromy structure of each brane.
Since this task is beyond the scope of this work, 
we would like to study this issue 
in the framework of supergravity theories for nongeometric backgrounds \cite{Andriot:2013xca, Andriot:2014uda, Blair:2014zba, Sakatani:2014hba}
in future work.

\section{Conclusion and discussions}
\label{sect:discussions}

In this paper, we have studied the supersymmetry projection rules on various exotic branes in type II string theories and M-theory.
Following the string dualities acting on the mass formulae and the supersymmetry parameters, we obtained explicit expressions for the projection rules.
By virtue of these rules, we discussed the exotic duality among the defect branes in type II theories.
Furthermore, we 
considered the mapping from
a configuration of multiple non-parallel standard branes
to that of multiple non-parallel exotic branes.
Applying the exotic 
mapping
to the configurations in which a brane is ending on another brane, we could read off the situations 
in which (exotic) branes can be ending on (exotic) branes.
Although this is analogous to the case of the standard branes,
we should stress that the latter configuration cannot be obtained from the former via the string dualities. This is because the latter configuration is only realized in nongeometric backgrounds in string theory. This should be analyzed in terms of the extended versions of supergravity theories, called
$\beta$-supergravity or $\gamma$-supergravity.

In this work we have not seriously considered
the global structure of the spacetime modified by the non-trivial monodromy caused by the string dualities, and the back reactions originating from the strong tensions of the exotic branes.
However, we can trust the supersymmetry projection rules on the exotic branes as far as they concern the supersymmetric configurations.

In the configuration in which
the exotic $1^6_4$-brane is ending on the exotic $p^{7-p}_3$-brane, 
we would expect that the excitations on the $p^{7-p}_3$-branes could be evaluated in terms of the mode expansions of the exotic $1^6_4$-brane in the string coupling region $g_s \to \infty$.
This is a naive analogy from the configuration that the F-string is ending on the D-brane.
In order to analyze this issue,
extended supergravity theories such as $\beta$-supergravity and its further extension which contains the dualized Ramond-Ramond potentials might be a strong framework.

\section*{Acknowledgements}

The author thanks
Tetsutaro Higaki,
Yosuke Imamura,
Hirotaka Kato,
Shin Sasaki,
Masaki Shigemori
and 
Hongfei Shu
for helpful discussions and comments.
He also thanks the Yukawa Institute for Theoretical Physics at Kyoto University for hospitality during the YITP workshop on ``Microstructures of black holes'' ({YITP-W-15-20}).
This work is supported by the MEXT-Supported Program for the Strategic Research Foundation at Private Universities ``Topological Science'' ({Grant No.~S1511006}). 
It is also supported in part by the Iwanami-Fujukai Foundation.

\begin{appendix}
\section*{Appendix}

\section{Computations}
\label{app:data}

In this appendix we exhibit the supersymmetry projection rules on various exotic branes derived from the string dualities (\ref{string-dualities}), (\ref{TS-rules-SUSY}) and (\ref{mass-IIA2M}). 

\subsection{SUSY projection rules on solitonic branes}

\subsubsection*{Descendants from the IIA NS5-brane}

\begin{itemize}
\item $\text{IIA NS5(12345)} \xrightarrow{\text{T}_9} \text{IIB KK5(12345,9)}$:
\begin{alignat}{2}
\text{IIA NS5(12345)} : && \ \ \ 
\pm \eps \ &= \ 
\Gamma^{012345} \eps
\, , \nn \\
&& \text{i.e.,} \ \ 
\pm \eps_L \ &= \ 
\Gamma^{012345} \eps_L
\, , \ls
\pm \eps_R \ = \ 
\Gamma^{012345} \eps_R
\, , \nn \\
\text{T$_9$-duality}: && \ \ \ 
\eps_R \ &\to \ \Gamma^9 \Gamma \eps_R
\, , \nn \\
&& 
\pm \eps_R \ &= \ 
\Gamma \Gamma^9 \Gamma^{012345} \Gamma^9 \Gamma \eps_R
\ = \ 
\Gamma^{012345} \eps_R
\, , \nn \\
\therefore \ \ \ 
\text{IIB KK5(12345,9)} : && \ \ \ 
\pm \eps_L \ &= \ 
\Gamma^{012345} \eps_L
\, , \ls
\pm \eps_R \ = \ 
\Gamma^{012345} \eps_R
\, , \nn \\
&& \text{namely,} \ \ \ 
\pm \eps \ &= \ 
\Gamma^{012345} \eps
\, .
\end{alignat}

\item $\text{IIB KK5(12345,9)} \xrightarrow{\text{T}_8} \text{IIA $5^2_2(12345,89)$}$:
\begin{alignat}{2}
\text{IIB KK5(12345,9)} : && \ \ \ 
\pm \eps \ &= \ 
\Gamma^{012345} \eps
\, , \nn \\
&& \text{i.e.,} \ \ 
\pm \eps_L \ &= \ 
\Gamma^{012345} \eps_L
\, , \ls
\pm \eps_R \ = \ 
\Gamma^{012345} \eps_R
\, , \nn \\
\text{T$_8$-duality}: && \ \ \ 
\eps_R \ &\to \ \Gamma^8 \Gamma \eps_R
\, , \nn \\
&&
\pm \eps_R \ &= \ 
\Gamma \Gamma^8 \Gamma^{012345} \Gamma^8 \Gamma \eps_R
\ = \ 
\Gamma^{012345} \eps_R
\, , \nn \\
\therefore \ \ \ 
\text{IIA $5^2_2(12345,89)$} : && \ \ \ 
\pm \eps_L \ &= \ 
\Gamma^{012345} \eps_L
\, , \ls
\pm \eps_R \ = \ 
\Gamma^{012345} \eps_R
\, , \nn \\
&& \text{namely,} \ \ \ 
\pm \eps \ &= \ 
\Gamma^{012345} \eps
\, .
\end{alignat}

\item $\text{IIA $5^2_2(12345,89)$} \xrightarrow{\text{T}_7} \text{IIB $5^3_2(12345,789)$}$:
\begin{alignat}{2}
\text{IIA $5^2_2(12345,89)$} : && \ \ \ 
\pm \eps \ &= \ 
\Gamma^{012345} \eps
\, , \nn \\
&& \text{i.e.,} \ \ 
\pm \eps_L \ &= \ 
\Gamma^{012345} \eps_L
\, , \ls
\pm \eps_R \ = \ 
\Gamma^{012345} \eps_R
\, , \nn \\
\text{T$_7$-duality}: && \ \ \ 
\eps_R \ &\to \ \Gamma^7 \Gamma \eps_R
\, , \nn \\
&&
\pm \eps_R \ &= \ 
\Gamma \Gamma^7 \Gamma^{012345} \Gamma^7 \Gamma \eps_R
\ = \ 
\Gamma^{012345} \eps_R
\, , \nn \\
\therefore \ \ \ 
\text{IIB $5^3_2(12345,789)$} : && \ \ \ 
\pm \eps_L \ &= \ 
\Gamma^{012345} \eps_L
\, , \ls
\pm \eps_R \ = \ 
\Gamma^{012345} \eps_R
\, , \nn \\
&& \text{namely,} \ \ \ 
\pm \eps \ &= \ 
\Gamma^{012345} \eps
\, .
\end{alignat}

\item $\text{IIB $5^3_2(12345,789)$} \xrightarrow{\text{T}_6} \text{IIA $5^4_2(12345,6789)$}$:
\begin{alignat}{2}
\text{IIB $5^3_2(12345,789)$} : && \ \ \ 
\pm \eps \ &= \ 
\Gamma^{012345} \eps
\, , \nn \\
&& \text{i.e.,} \ \ 
\pm \eps_L \ &= \ 
\Gamma^{012345} \eps_L
\, , \ls
\pm \eps_R \ = \ 
\Gamma^{012345} \eps_R
\, , \nn \\
\text{T$_6$-duality}: && \ \ \
\eps_R \ &\to \ \Gamma^6 \Gamma \eps_R
\, , \nn \\
&&
\pm \eps_R \ &= \ 
\Gamma \Gamma^6 \Gamma^{012345} \Gamma^6 \Gamma \eps_R
\ = \ 
\Gamma^{012345} \eps_R
\, , \nn \\
\therefore \ \ \ 
\text{IIA $5^4_2(12345,6789)$} : && \ \ \ 
\pm \eps_L \ &= \ 
\Gamma^{012345} \eps_L
\, , \ls
\pm \eps_R \ = \ 
\Gamma^{012345} \eps_R
\, , \nn \\
&& \text{namely,} \ \ \ 
\pm \eps \ &= \ 
\Gamma^{012345} \eps
\, .
\end{alignat}

\end{itemize}

\subsubsection*{Descendants from the IIB NS5-brane}

\begin{itemize}

\item $\text{IIB NS5(12345)} \xrightarrow{\text{T}_9} \text{IIA KK5(12345,9)}$:
\begin{alignat}{2}
\text{IIB NS5(12345)} : && \ \ \ 
\pm \eps \ &= \ 
\Gamma^{012345} (\sigma_3) \eps
\, , \nn \\
&& \text{i.e.,} \ \ 
\pm \eps_L \ &= \ 
\Gamma^{012345} \eps_L
\, , \ls
\pm \eps_R \ = \ 
- \Gamma^{012345} \eps_R
\, , \nn \\
\text{T$_9$-duality}: && \ \ \  
\eps_R \ &\to \ \Gamma^9 \Gamma \eps_R
\, , \nn \\
&&
\pm \eps_R \ &= \ 
- \Gamma \Gamma^9 \Gamma^{012345} \Gamma^9 \Gamma \eps_R
\ = \ 
- \Gamma^{012345} \eps_R
\, , \nn \\
\therefore \ \ \ 
\text{IIA KK5(12345,9)} : && \ \ \ 
\pm \eps_L \ &= \ 
\Gamma^{012345} \eps_L
\, , \ls
\pm \eps_R \ = \ 
- \Gamma^{012345} \eps_R
\, , \nn \\
&& \text{namely,} \ \ \ 
\pm \eps \ &= \ 
\Gamma^{012345} \Gamma \eps
\, .
\end{alignat}

\item $\text{IIA KK5(12345,9)} \xrightarrow{\text{T}_8} \text{IIB $5^2_2(12345,89)$}$:
\begin{alignat}{2}
\text{IIA KK5(12345,9)} : && \ \ \ 
\pm \eps \ &= \ 
\Gamma^{012345} \Gamma \eps
\, , \nn \\
&& \text{i.e.,} \ \ 
\pm \eps_L \ &= \ 
\Gamma^{012345} \eps_L
\, , \ls
\pm \eps_R \ = \ 
- \Gamma^{012345} \eps_R
\, , \nn \\
\text{T$_8$-duality}: && \ \ \  
\eps_R \ &\to \ \Gamma^8 \Gamma \eps_R
\, , \nn \\
&&
\pm \eps_R \ &= \ 
- \Gamma \Gamma^8 \Gamma^{012345} \Gamma^8 \Gamma \eps_R
\ = \ 
- \Gamma^{012345} \eps_R
\, , \nn \\
\therefore \ \ \ 
\text{IIB $5^2_2(12345,89)$} : && \ \ \ 
\pm \eps_L \ &= \ 
\Gamma^{012345} \eps_L
\, , \ls
\pm \eps_R \ = \ 
- \Gamma^{012345} \eps_R
\, , \nn \\
&& \text{namely,} \ \ \ 
\pm \eps \ &= \ 
\Gamma^{012345} (\sigma_3) \eps
\, .
\end{alignat}

\item $\text{IIB $5^2_2(12345,89)$} \xrightarrow{\text{T}_7} \text{IIA $5^3_2(12345,789)$}$:
\begin{alignat}{2}
\text{IIB $5^2_2(12345,89)$} : && \ \ \ 
\pm \eps \ &= \ 
\Gamma^{012345} (\sigma_3) \eps
\, , \nn \\
&& \text{i.e.,} \ \ 
\pm \eps_L \ &= \ 
\Gamma^{012345} \eps_L
\, , \ls
\pm \eps_R \ = \ 
- \Gamma^{012345} \eps_R
\, , \nn \\
\text{T$_7$-duality}: && \ \ \
\eps_R \ &\to \ \Gamma^7 \Gamma \eps_R
\, , \nn \\
&&
\pm \eps_R \ &= \ 
- \Gamma \Gamma^7 \Gamma^{012345} \Gamma^7 \Gamma \eps_R
\ = \ 
- \Gamma^{012345} \eps_R
\, , \nn \\
\therefore \ \ \ 
\text{IIA $5^3_2(12345,789)$} : && \ \ \ 
\pm \eps_L \ &= \ 
\Gamma^{012345} \eps_L
\, , \ls
\pm \eps_R \ = \ 
- \Gamma^{012345} \eps_R
\, , \nn \\
&& \text{namely,} \ \ \ 
\pm \eps \ &= \ 
\Gamma^{012345} \Gamma \eps
\, .
\end{alignat}

\item $\text{IIA $5^3_2(12345,789)$} \xrightarrow{\text{T}_6} \text{IIB $5^4_2(12345,6789)$}$:
\begin{alignat}{2}
\text{IIA $5^3_2(12345,789)$} : && \ \ \ 
\pm \eps \ &= \ 
\Gamma^{012345} \Gamma \eps
\, , \nn \\
&& \text{i.e.,} \ \ 
\pm \eps_L \ &= \ 
\Gamma^{012345} \eps_L
\, , \ls
\pm \eps_R \ = \ 
- \Gamma^{012345} \eps_R
\, , \nn \\
\text{T$_6$-duality}: && \ \ \
\eps_R \ &\to \ \Gamma^6 \Gamma \eps_R
\, , \nn \\
&&
\pm \eps_R \ &= \ 
- \Gamma \Gamma^8 \Gamma^{012345} \Gamma^8 \Gamma \eps_R
\ = \ 
- \Gamma^{012345} \eps_R
\, , \nn \\
\therefore \ \ \ 
\text{IIB $5^4_2(12345,6789)$} : && \ \ \ 
\pm \eps_L \ &= \ 
\Gamma^{012345} \eps_L
\, , \ls
\pm \eps_R \ = \ 
- \Gamma^{012345} \eps_R
\, , \nn \\
&& \text{namely,} \ \ \ 
\pm \eps \ &= \ 
\Gamma^{012345} (\sigma_3) \eps
\, .
\end{alignat}

\end{itemize}

\subsection{SUSY projection rules on defect branes}

\subsubsection*{Descendants from the IIB D7-brane}

\begin{itemize}

\item $\text{IIB D7(1234567)} \xrightarrow{\text{S}} \text{IIB $7_3(1234567)$}$:
\begin{alignat}{2}
\text{D7(1234567)}: && \ \ \ 
\pm \eps \ &= \ 
\Gamma^{01234567} (\I \sigma_2) \eps
\, , \nn \\
\text{S-duality}: && \ \ \
\eps \ &\to \ S \eps
\, , \nn \\
\therefore \ \ \ 
\text{IIB $7_3(1234567)$} : && \ \ \ 
\pm \eps \ &= \ 
\Gamma^{01234567} S^{-1} (\I \sigma_2) S \eps
\nn \\
&&&= \ 
\Gamma^{01234567} (\I \sigma_2) \eps
\, . 
\end{alignat}

\item $\text{IIB $7_3(1234567)$} \xrightarrow{\text{T}_7} \text{IIA $6^1_3 (123456,7)$}$: 
\begin{alignat}{2}
\text{IIB $7_3(1234567)$} : && \ \ \ 
\pm \eps \ &= \ 
\Gamma^{01234567} (\I \sigma_2) \eps
\, , \nn \\
&& \text{i.e.,} \ \ 
\pm \eps_L \ &= \ 
\Gamma^{01234567} \eps_R
\, , \nn \\
\text{T$_7$-duality}: && \ \ \
\eps_R \ &\to \ \Gamma^7 \Gamma \eps_R
\, , \nn \\
&& \ \ \ 
\pm \eps_L \ &= \ 
\Gamma^{01234567} \Gamma^7 \Gamma \eps_R
\ = \ 
- \Gamma^{0123456} \eps_R
\, , \nn \\
\therefore \ \ \ 
\text{IIA $6^1_3(123456,7)$} : && \ \ \ 
\pm \eps_L \ &= \ 
- \Gamma^{0123456} \eps_R
\, , \nn \\
&& \text{namely,} \ \ \ 
\pm \eps \ &= \ 
- \Gamma^{0123456} (\sigma_1) \eps_R
\, .
\end{alignat}
On the right-hand side, there is a negative sign that should not distract readers
because this merely comes from the anti-commutation relations among the Dirac gamma matrices.
In later computations we often encounter the same situation.

\item $\text{IIA $6^1_3(123456,7)$} \xrightarrow{\text{T}_6} \text{IIB $5^2_3 (12345,67)$}$: 
\begin{alignat}{2}
\text{IIA $6^1_3(123456,7)$} : && \ \ \ 
\pm \eps \ &= \ 
- \Gamma^{0123456} (\sigma_1) \eps
\, , \nn \\
&& \text{i.e.,} \ \ 
\pm \eps_L \ &= \ 
- \Gamma^{0123456} \eps_R
\, , \nn \\
\text{T$_6$-duality}: && \ \ \ 
\eps_R \ &\to \ \Gamma^6 \Gamma \eps_R
\, , \nn \\
&&
\pm \eps_L \ &= \ 
- \Gamma^{0123456} \Gamma^6 \Gamma \eps_R
\ = \ 
- \Gamma^{012345} \eps_R
\, , \nn \\
\therefore \ \ \ 
\text{IIB $5^2_3(12345,67)$} : && \ \ \ 
\pm \eps_L \ &= \ 
- \Gamma^{012345} \eps_R
\, , \nn \\
&& \text{namely,} \ \ \ 
\pm \eps \ &= \ 
- \Gamma^{012345} (\sigma_1) \eps
\, .
\end{alignat}

\item $\text{IIB $5^2_3(12345,67)$} \xrightarrow{\text{T}_5} \text{IIA $4^3_3 (1234,567)$}$: 
\begin{alignat}{2}
\text{IIB $5^2_3(12345,67)$} : && \ \ \ 
\pm \eps \ &= \ 
- \Gamma^{012345} (\sigma_1) \eps
\, , \nn \\
&& \text{i.e.,} \ \ 
\pm \eps_L \ &= \ 
- \Gamma^{012345} \eps_R
\, , \nn \\
\text{T$_5$-duality}: && \ \ \
\eps_R \ &\to \ \Gamma^5 \Gamma \eps_R
\, , \nn \\
&&
\pm \eps_L \ &= \ 
- \Gamma^{012345} \Gamma^5 \Gamma \eps_R
\ = \ 
+ \Gamma^{01234} \eps_R
\, , \nn \\
\therefore \ \ \ 
\text{IIA $4^3_3(1234,567)$} : && \ \ \ 
\pm \eps_L \ &= \ 
\Gamma^{01234} \eps_R
\, , \nn \\
&& \text{namely,} \ \ \ 
\pm \eps \ &= \ 
\Gamma^{01234} \Gamma (\sigma_1) \eps
\, .
\end{alignat}

\item $\text{IIA $4^3_3(1234,567)$} \xrightarrow{\text{T}_4} \text{IIB $3^4_3 (123,4567)$}$: 
\begin{alignat}{2}
\text{IIA $4^3_3(1234,567)$} : && \ \ \ 
\pm \eps \ &= \ 
\Gamma^{01234} \Gamma (\sigma_1) \eps
\, , \nn \\
&& \text{i.e.,} \ \ 
\pm \eps_L \ &= \ 
\Gamma^{01234} \eps_R
\, , \nn \\
\text{T$_4$-duality}: && \ \ \
\eps_R \ &\to \ \Gamma^4 \Gamma \eps_R
\, , \nn \\
&&
\pm \eps_L \ &= \ 
\Gamma^{01234} \Gamma^4 \Gamma \eps_R
\ = \
+ \Gamma^{0123} \eps_R
\, , \nn \\
\therefore \ \ \ 
\text{IIB $3^4_3(123,4567)$} : && \ \ \ 
\pm \eps_L \ &= \ 
\Gamma^{0123} \eps_R
\, , \nn \\
&& \text{namely,} \ \ \ 
\pm \eps \ &= \ 
\Gamma^{0123} (\I \sigma_2) \eps
\, .
\end{alignat}

\item $\text{IIB $3^4_3(123,4567)$} \xrightarrow{\text{T}_3} \text{IIA $2^5_3 (12,34567)$}$: 
\begin{alignat}{2}
\text{IIB $3^4_3(123,4567)$} : && \ \ \ 
\pm \eps \ &= \ 
\Gamma^{0123} (\I \sigma_2) \eps
\, , \nn \\
&& \text{i.e.,} \ \ 
\pm \eps_L \ &= \ 
\Gamma^{0123} \eps_R
\, , \nn \\
\text{T$_3$-duality}: && \ \ \
\eps_R \ &\to \ \Gamma^3 \Gamma \eps_R
\, , \nn \\
&&
\pm \eps_L \ &= \ 
\Gamma^{0123} \Gamma^3 \Gamma \eps_R
\ = \ 
- \Gamma^{012} \eps_R
\, , \nn \\
\therefore \ \ \ 
\text{IIA $2^5_3(12,34567)$} : && \ \ \ 
\pm \eps_L \ &= \ 
- \Gamma^{012} \eps_R
\, , \nn \\
&& \text{namely,} \ \ \ 
\pm \eps \ &= \ 
- \Gamma^{012} (\sigma_1) \eps
\, .
\end{alignat}

\item $\text{IIA $2^5_3(12,34567)$} \xrightarrow{\text{T}_2} \text{IIB $1^6_3 (1,234567)$}$: 
\begin{alignat}{2}
\text{IIA $2^5_3(12,34567)$} : && \ \ \ 
\pm \eps \ &= \ 
- \Gamma^{012} (\sigma_1) \eps
\, , \nn \\
&& \text{i.e.,} \ \ 
\pm \eps_L \ &= \ 
- \Gamma^{012} \eps_R
\, , \nn \\
\text{T$_2$-duality}: && \ \ \
\eps_R \ &\to \ \Gamma^2 \Gamma \eps_R
\, , \nn \\
&&
\pm \eps_L \ &= \ 
- \Gamma^{012} \Gamma^2 \Gamma \eps_R
\ = \ 
- \Gamma^{01} \eps_R
\, , \nn \\
\therefore \ \ \ 
\text{IIB $1^6_3(1,234567)$} : && \ \ \ 
\pm \eps_L \ &= \ 
- \Gamma^{01} \eps_R
\, , \nn \\
&& \text{namely,} \ \ \ 
\pm \eps \ &= \ 
- \Gamma^{01} (\sigma_1) \eps
\, .
\end{alignat}

\item $\text{IIB $1^6_3(1,234567)$} \xrightarrow{\text{T}_1} \text{IIA $0^7_3 (,1234567)$}$: 
\begin{alignat}{2}
\text{IIB $1^6_3(1,234567)$} : && \ \ \ 
\pm \eps \ &= \ 
- \Gamma^{01} (\sigma_1) \eps
\, , \nn \\
&& \text{i.e.,} \ \ 
\pm \eps_L \ &= \ 
- \Gamma^{01} \eps_R
\, , \nn \\
\text{T$_1$-duality}: && \ \ \
\eps_R \ &\to \ \Gamma^1 \Gamma \eps_R
\, , \nn \\
&&
\pm \eps_L \ &= \ 
- \Gamma^{01} \Gamma^1 \Gamma \eps_R
\ = \ 
+ \Gamma^{0} \eps_R
\, , \nn \\
\therefore \ \ \ 
\text{IIA $0^7_3(,1234567)$} : && \ \ \ 
\pm \eps_L \ &= \ 
\Gamma^{0} \eps_R
\, , \nn \\
&& \text{namely,} \ \ \ 
\pm \eps \ &= \ 
\Gamma^{0} \Gamma (\sigma_1) \eps
\, .
\end{alignat}

\end{itemize}

\subsubsection*{Descendants from IIB $p^{7-p}_3$-branes via S-duality}

\begin{itemize}
\item $\text{IIB $5^2_3(12345,67)$} \xrightarrow{\text{S}} \text{IIB $5^2_2 (12345,67)$}$: 
\begin{alignat}{2}
\text{IIB $5^2_3(12345,67)$}: && \ \ \ 
\pm \eps \ &= \ 
- \Gamma^{012345} (\sigma_1) \eps
\, , \nn \\
\text{S-duality}: && \ \ \
\eps \ &\to \ S \eps
\, , \nn \\
\therefore \ \ \ 
\text{IIB $5^2_2(12345,67)$} : && \ \ \ 
\pm \eps \ &= \ 
- \Gamma^{012345} S^{-1} (\sigma_1) S \eps
\nn \\
&&&= \ 
- \Gamma^{012345} (\sigma_3) \eps
\, . 
\end{alignat}

\item $\text{IIB $3^4_3(1234,567)$} \xrightarrow{\text{S}} \text{IIB $3^4_3 (123,4567)$}$: 
\begin{alignat}{2}
\text{IIB $3^4_3(123,4567)$}: && \ \ \ 
\pm \eps \ &= \ 
\Gamma^{0123} (\I \sigma_2) \eps
\, , \nn \\
\text{S-duality}: && \ \ \
\eps \ &\to \ S \eps
\, , \nn \\
\therefore \ \ \ 
\text{IIB $3^4_3(123,4567)$} : && \ \ \ 
\pm \eps \ &= \ 
\Gamma^{0123} S^{-1} (\I \sigma_2) S \eps
\nn \\
&&&= \ 
\Gamma^{0123} (\I \sigma_2) \eps
\ls 
\text{(self-dual)}
\, . 
\end{alignat}

\item $\text{IIB $1^6_3(1,234567)$} \xrightarrow{\text{S}} \text{IIB $1^6_4 (1,234567)$}$: 
\begin{alignat}{2}
\text{IIB $1^6_3(1,234567)$}: && \ \ \ 
\pm \eps \ &= \ 
- \Gamma^{01} (\sigma_1) \eps
\, , \nn \\
\text{S-duality}: && \ \ \
\eps \ &\to \ S \eps
\, , \nn \\
\therefore \ \ \ 
\text{IIB $1^6_4(1,234567)$} : && \ \ \ 
\pm \eps \ &= \ 
- \Gamma^{01} S^{-1} (\sigma_1) S \eps
\nn \\
&&&= \ 
- \Gamma^{01} (\sigma_3) \eps
\, . 
\end{alignat}
\end{itemize}

\subsubsection*{Descendants from the IIB $1^6_4$-brane via T-duality}

\begin{itemize}
\item $\text{IIB $1^6_4(1,234567)$} \xrightarrow{\text{T}_2} \text{IIA $1^6_4 (1,234567)$}$: 
\begin{alignat}{2}
\text{IIB $1^6_4(1,234567)$} : && \ \ \ 
\pm \eps \ &= \ 
- \Gamma^{01} (\sigma_3) \eps
\, , \nn \\
&& \text{i.e.,} \ \ 
\pm \eps_L \ &= \ 
- \Gamma^{01} \eps_L
\, , \ls
\pm \eps_R \ = \ 
+ \Gamma^{01} \eps_R
\, , \nn \\
\text{T$_2$-duality}: && \ \ \
\eps_R \ &\to \ \Gamma^2 \Gamma \eps_R
\, , \nn \\
&&
\pm \eps_R \ &= \ 
+ \Gamma \Gamma^2 \Gamma^{01} \Gamma^2 \Gamma \eps_R
\ = \ 
+ \Gamma^{01} \eps_R
\, , \nn \\
\therefore \ \ \ 
\text{IIA $1^6_4(1,234567)$} : && \ \ \ 
\pm \eps_L \ &= \ 
- \Gamma^{01} \eps_L
\, , \ls
\pm \eps_R \ = \ 
+ \Gamma^{01} \eps_R
\, , \nn \\
&& \text{namely,} \ \ \ 
\pm \eps \ &= \ 
- \Gamma^{01} \Gamma \eps
\, .
\end{alignat}

\item $\text{IIB $1^6_4(1,234567)$} \xrightarrow{\text{T}_1} \text{IIA $0^{(1,6)}_4 (,234567,1)$}$: 
\begin{alignat}{2}
\text{IIB $1^6_4(1,234567)$} : && \ \ \ 
\pm \eps \ &= \ 
- \Gamma^{01} (\sigma_3) \eps
\, , \nn \\
&& \text{i.e.,} \ \ 
\pm \eps_L \ &= \ 
- \Gamma^{01} \eps_L
\, , \ls
\pm \eps_R \ = \ 
+ \Gamma^{01} \eps_R
\, , \nn \\
\text{T$_1$-duality}: && \ \ \
\eps_R \ &\to \ \Gamma^1 \Gamma \eps_R
\, , \nn \\
&&
\pm \eps_R \ &= \ 
+ \Gamma \Gamma^1 \Gamma^{01} \Gamma^1 \Gamma \eps_R
\ = \ 
- \Gamma^{01} \eps_R
\, , \nn \\
\therefore \ \ \ 
\text{IIA $0^{(1,6)}_4(,234567,1)$} : && \ \ \ 
\pm \eps_L \ &= \ 
- \Gamma^{01} \eps_L
\, , \ls
\pm \eps_R \ = \ 
- \Gamma^{01} \eps_R
\, , \nn \\
&& \text{namely,} \ \ \ 
\pm \eps \ &= \ 
- \Gamma^{01} \eps
\, .
\end{alignat}

\item $\text{IIA $1^6_4(1,234567)$} \xrightarrow{\text{T}_1} \text{IIB $0^{(1,6)}_4 (,234567,1)$}$: 
\begin{alignat}{2}
\text{IIA $1^6_4(1,234567)$} : && \ \ \ 
\pm \eps \ &= \ 
- \Gamma^{01} \Gamma \eps
\, , \nn \\
&& \text{i.e.,} \ \ 
\pm \eps_L \ &= \ 
- \Gamma^{01} \eps_L
\, , \ls
\pm \eps_R \ = \ 
+ \Gamma^{01} \eps_R
\, , \nn \\
\text{T$_1$-duality}: && \ \ \
\eps_R \ &\to \ \Gamma^1 \Gamma \eps_R
\, , \nn \\
&&
\pm \eps_R \ &= \ 
+ \Gamma \Gamma^1 \Gamma^{01} \Gamma^1 \Gamma \eps_R
\ = \ 
- \Gamma^{01} \eps_R
\, , \nn \\
\therefore \ \ \ 
\text{IIB $0^{(1,6)}_4(,234567,1)$} : && \ \ \ 
\pm \eps_L \ &= \ 
- \Gamma^{01} \eps_L
\, , \ls
\pm \eps_R \ = \ 
- \Gamma^{01} \eps_R
\, , \nn \\
&& \text{namely,} \ \ \ 
\pm \eps \ &= \ 
- \Gamma^{01} \eps
\, .
\end{alignat}

\end{itemize}

\subsection{SUSY projection rules on domain walls}

\subsubsection*{Domain walls from defect branes via T-duality}

\begin{itemize}
\item $\text{IIB $7_3(1234567)$} \xrightarrow{\text{T}_9} \text{IIA $7^{(1,0)}_3 (1234567,,9)$}$: 
\begin{alignat}{2}
\text{IIB $7_3(1234567)$} : && \ \ \ 
\pm \eps \ &= \ 
\Gamma^{01234567} (\I \sigma_2) \eps
\, , \nn \\
&& \text{i.e.,} \ \ 
\pm \eps_L \ &= \ 
\Gamma^{01234567} \eps_R
\, , \nn \\
\text{T$_9$-duality}: && \ \ \
\eps_R \ &\to \ \Gamma^9 \Gamma \eps_R
\, , \nn \\
&&
\pm \eps_L \ &= \ 
\Gamma^{01234567} \Gamma^9 \Gamma \eps_R
\ = \ 
- \Gamma^{012345679} \eps_R
\, , \nn \\
\therefore \ \ \ 
\text{IIA $7^{(1,0)}_3(1234567,,9)$} : && \ \ \ 
\pm \eps_L \ &= \ 
- \Gamma^{012345679} \eps_R
\, , \nn \\
&& \text{namely,} \ \ \ 
\pm \eps \ &= \ 
- \Gamma^{012345679} \Gamma (\sigma_1) \eps_R
\, .
\end{alignat}

\item $\text{IIA $6^1_3(123456,7)$} \xrightarrow{\text{T}_9} \text{IIB $6^{(1,1)}_3 (123456,7,9)$}$: 
\begin{alignat}{2}
\text{IIA $6^1_3(123456,7)$} : && \ \ \ 
\pm \eps \ &= \ 
- \Gamma^{0123456} (\sigma_1) \eps
\, , \nn \\
&& \text{i.e.,} \ \ 
\pm \eps_L \ &= \ 
- \Gamma^{0123456} \eps_R
\, , \nn \\
\text{T$_9$-duality}: && \ \ \
\eps_R \ &\to \ \Gamma^9 \Gamma \eps_R
\, , \nn \\
&&
\pm \eps_L \ &= \ 
- \Gamma^{0123456} \Gamma^9 \Gamma \eps_R
\ = \ 
- \Gamma^{01234569} \eps_R
\, , \nn \\
\therefore \ \ \ 
\text{IIB $6^{(1,1)}_3(123456,7,9)$} : && \ \ \ 
\pm \eps_L \ &= \ 
- \Gamma^{01234569} \eps_R
\, , \nn \\
&& \text{namely,} \ \ \ 
\pm \eps \ &= \ 
- \Gamma^{01234569} (\I \sigma_2) \eps
\, .
\end{alignat}

\item $\text{IIB $5^2_3(12345,67)$} \xrightarrow{\text{T}_9} \text{IIA $5^{(1,2)}_3 (12345,67,9)$}$: 
\begin{alignat}{2}
\text{IIB $5^2_3(12345,67)$} : && \ \ \ 
\pm \eps \ &= \ 
- \Gamma^{012345} (\sigma_1) \eps
\, , \nn \\
&& \text{i.e.,} \ \ 
\pm \eps_L \ &= \ 
- \Gamma^{012345} \eps_R
\, , \nn \\
\text{T$_9$-duality}: && \ \ \
\eps_R \ &\to \ \Gamma^9 \Gamma \eps_R
\, , \nn \\
&&
\pm \eps_L \ &= \ 
- \Gamma^{012345} \Gamma^9 \Gamma \eps_R
\ = \ 
+ \Gamma^{0123459} \eps_R
\, , \nn \\
\therefore \ \ \ 
\text{IIA $5^{(1,2)}_3(12345,67,9)$} : && \ \ \ 
\pm \eps_L \ &= \ 
\Gamma^{0123459} \eps_R
\, , \nn \\
&& \text{namely,} \ \ \ 
\pm \eps \ &= \ 
\Gamma^{0123459} (\sigma_1) \eps
\, .
\end{alignat}

\item $\text{IIA $4^3_3(1234,567)$} \xrightarrow{\text{T}_9} \text{IIB $4^{(1,3)}_3 (1234,567,9)$}$: 
\begin{alignat}{2}
\text{IIA $4^3_3(1234,567)$} : && \ \ \ 
\pm \eps \ &= \ 
\Gamma^{01234} (\I \sigma_2) \eps
\, , \nn \\
&& \text{i.e.,} \ \ 
\pm \eps_L \ &= \ 
\Gamma^{01234} \eps_R
\, , \nn \\
\text{T$_9$-duality}: && \ \ \
\eps_R \ &\to \ \Gamma^9 \Gamma \eps_R
\, , \nn \\
&&
\pm \eps_L \ &= \ 
\Gamma^{01234} \Gamma^9 \Gamma \eps_R
\ = \ 
+ \Gamma^{012349} \eps_R
\, , \nn \\
\therefore \ \ \ 
\text{IIB $4^{(1,3)}_3(1234,567,9)$} : && \ \ \ 
\pm \eps_L \ &= \ 
\Gamma^{012349} \eps_R
\, , \nn \\
&& \text{namely,} \ \ \ 
\pm \eps \ &= \ 
\Gamma^{012349} (\sigma_1) \eps
\, .
\end{alignat}

\item $\text{IIB $3^4_3(123,4567)$} \xrightarrow{\text{T}_9} \text{IIA $3^{(1,4)}_3 (123,4567,9)$}$: 
\begin{alignat}{2}
\text{IIB $3^4_3(123,4567)$} : && \ \ \ 
\pm \eps \ &= \ 
\Gamma^{0123} (\I \sigma_2) \eps
\, , \nn \\
&& \text{i.e.,} \ \ 
\pm \eps_L \ &= \ 
\Gamma^{0123} \eps_R
\, , \nn \\
\text{T$_9$-duality}: && \ \ \
\eps_R \ &\to \ \Gamma^9 \Gamma \eps_R
\, , \nn \\
&& 
\pm \eps_L \ &= \ 
\Gamma^{0123} \Gamma^9 \Gamma \eps_R
\ = \ 
- \Gamma^{01239} \eps_R
\, , \nn \\
\therefore \ \ \ 
\text{IIA $3^{(1,4)}_3(123,4567,9)$} : && \ \ \ 
\pm \eps_L \ &= \ 
- \Gamma^{01239} \eps_R
\, , \nn \\
&& \text{namely,} \ \ \ 
\pm \eps \ &= \ 
- \Gamma^{01239} \Gamma (\sigma_1) \eps
\, .
\end{alignat}

\item $\text{IIA $2^5_3(12,34567)$} \xrightarrow{\text{T}_9} \text{IIB $2^{(1,5)}_3 (12,34567,9)$}$: 
\begin{alignat}{2}
\text{IIA $2^5_3(12,34567)$} : && \ \ \ 
\pm \eps \ &= \ 
- \Gamma^{012} (\sigma_1) \eps
\, , \nn \\
&& \text{i.e.,} \ \ 
\pm \eps_L \ &= \ 
- \Gamma^{012} \eps_R
\, , \nn \\
\text{T$_9$-duality}: && \ \ \
\eps_R \ &\to \ \Gamma^9 \Gamma \eps_R
\, , \nn \\
&&
\pm \eps_L \ &= \ 
- \Gamma^{012} \Gamma^9 \Gamma \eps_R
\ = \ 
- \Gamma^{0129} \eps_R
\, , \nn \\
\therefore \ \ \ 
\text{IIB $2^{(1,5)}_3(12,34567,9)$} : && \ \ \ 
\pm \eps_L \ &= \ 
- \Gamma^{0129} \eps_R
\, , \nn \\
&& \text{namely,} \ \ \ 
\pm \eps \ &= \ 
- \Gamma^{0129} (\I \sigma_2) \eps
\, .
\end{alignat}

\item $\text{IIB $1^6_3(1,234567)$} \xrightarrow{\text{T}_9} \text{IIA $1^{(1,6)}_3 (1,234567,9)$}$: 
\begin{alignat}{2}
\text{IIB $1^6_3(1,234567)$} : && \ \ \ 
\pm \eps \ &= \ 
- \Gamma^{01} (\sigma_1) \eps
\, , \nn \\
&& \text{i.e.,} \ \ 
\pm \eps_L \ &= \ 
- \Gamma^{01} \eps_R
\, , \nn \\
\text{T$_9$-duality}: && \ \ \
\eps_R \ &\to \ \Gamma^9 \Gamma \eps_R
\, , \nn \\
&&
\pm \eps_L \ &= \ 
- \Gamma^{01} \Gamma^9 \Gamma \eps_R
\ = \ 
+ \Gamma^{019} \eps_R
\, , \nn \\
\therefore \ \ \ 
\text{IIA $1^{(1,6)}_3(1,234567,9)$} : && \ \ \ 
\pm \eps_L \ &= \ 
\Gamma^{019} \eps_R
\, , \nn \\
&& \text{namely,} \ \ \ 
\pm \eps \ &= \ 
\Gamma^{019} (\sigma_1) \eps
\, .
\end{alignat}

\item $\text{IIA $0^7_3(,1234567)$} \xrightarrow{\text{T}_9} \text{IIB $0^{(1,7)}_3 (,1234567,9)$}$: 
\begin{alignat}{2}
\text{IIA $0^7_3(,1234567)$} : && \ \ \ 
\pm \eps \ &= \ 
\Gamma^{0} \Gamma (\sigma_1) \eps
\, , \nn \\
&& \text{i.e.,} \ \ 
\pm \eps_L \ &= \ 
\Gamma^{0} \eps_R
\, , \nn \\
\text{T$_9$-duality}: && \ \ \
\eps_R \ &\to \ \Gamma^9 \Gamma \eps_R
\, , \nn \\
&&
\pm \eps_L \ &= \ 
\Gamma^{0} \Gamma^9 \Gamma \eps_R
\ = \
+ \Gamma^{09} \eps_R
\, , \nn \\
\therefore \ \ \ 
\text{IIB $0^{(1,7)}_3(,1234567,9)$} : && \ \ \ 
\pm \eps_L \ &= \ 
\Gamma^{09} \eps_R
\, , \nn \\
&& \text{namely,} \ \ \ 
\pm \eps \ &= \ 
\Gamma^{09} (\sigma_1) \eps
\, .
\end{alignat}

\end{itemize}

\subsubsection*{Descendants from the IIB $5^3_2$-brane}

\begin{itemize}

\item $\text{IIB $5^3_2(12345,789)$} \xrightarrow{\text{S}} \text{IIB $5^3_4(12345,789)$}$:
\begin{alignat}{2}
\text{IIB $5^3_2(12345,789)$}: && \ \ \ 
\pm \eps \ &= \ 
\Gamma^{012345} \eps
\, , \nn \\
\text{S-duality}: && \ \ \
\eps \ &\to \ S \eps
\, , \nn \\
\therefore \ \ \ 
\text{IIB $5^3_4(12345,789)$} : && \ \ \ 
\pm \eps \ &= \ 
\Gamma^{012345} S^{-1} S \eps
\nn \\
&&&= \ 
\Gamma^{012345} \eps
\, . 
\end{alignat}

\item $\text{IIB $5^3_4(12345,789)$} \xrightarrow{\text{T}_5} \text{IIA $4^{(1,3)}_4(1234,789,5)$}$:
\begin{alignat}{2}
\text{IIB $5^3_4(12345,789)$} : && \ \ \ 
\pm \eps \ &= \ 
\Gamma^{012345} \eps
\, , \nn \\
&& \text{i.e.,} \ \ 
\pm \eps_L \ &= \ 
\Gamma^{012345} \eps_L
\, , \ls
\pm \eps_R \ = \ 
\Gamma^{012345} \eps_R
\, , \nn \\
\text{T$_5$-duality}: && \ \ \
\eps_R \ &\to \ \Gamma^5 \Gamma \eps_R
\, , \nn \\
&&
\pm \eps_R \ &= \ 
\Gamma \Gamma^5 \Gamma^{012345} \Gamma^5 \Gamma \eps_R
\ = \ 
- \Gamma^{012345} \eps_R
\, , \nn \\
\therefore \ \ \ 
\text{IIA $4^{(1,3)}_4(1234,789,5)$} : && \ \ \ 
\pm \eps_L \ &= \ 
\Gamma^{012345} \eps_L
\, , \ls
\pm \eps_R \ = \ 
- \Gamma^{012345} \eps_R
\, , \nn \\
&& \text{namely,} \ \ \ 
\pm \eps \ &= \ 
\Gamma^{012345} \Gamma \eps
\, .
\end{alignat}

\item $\text{IIA $4^{(1,3)}_4(1234,789,5)$} \xrightarrow{\text{T}_4} \text{IIB $3^{(2,3)}_4(123,789,45)$}$:
\begin{alignat}{2}
\text{IIA $4^{(1,3)}_4(1234,789,5)$} : && \ \ \ 
\pm \eps \ &= \ 
\Gamma^{012345} \Gamma \eps
\, , \nn \\
&& \text{i.e.,} \ \ 
\pm \eps_L \ &= \ 
\Gamma^{012345} \eps_L
\, , \ls
\pm \eps_R \ = \ 
- \Gamma^{012345} \eps_R
\, , \nn \\
\text{T$_4$-duality}: && \ \ \
\eps_R \ &\to \ \Gamma^4 \Gamma \eps_R
\, , \nn \\
&&
\pm \eps_R \ &= \ 
- \Gamma \Gamma^4 \Gamma^{012345} \Gamma^4 \Gamma \eps_R
\ = \ 
\Gamma^{012345} \eps_R
\, , \nn \\
\therefore \ \ \ 
\text{IIB $3^{(2,3)}_4(123,789,45)$} : && \ \ \ 
\pm \eps_L \ &= \ 
\Gamma^{012345} \eps_L
\, , \ls
\pm \eps_R \ = \ 
\Gamma^{012345} \eps_R
\, , \nn \\
&& \text{namely,} \ \ \ 
\pm \eps \ &= \ 
\Gamma^{012345} \eps
\, .
\end{alignat}

\item $\text{IIB $3^{(2,3)}_4(123,789,45)$} \xrightarrow{\text{T}_3} \text{IIA $2^{(3,3)}_4(12,789,345)$}$:
\begin{alignat}{2}
\text{IIB $3^{(2,3)}_4(123,789,45)$} : && \ \ \ 
\pm \eps \ &= \ 
\Gamma^{012345} \eps
\, , \nn \\
&& \text{i.e.,} \ \ 
\pm \eps_L \ &= \ 
\Gamma^{012345} \eps_L
\, , \ls
\pm \eps_R \ = \ 
\Gamma^{012345} \eps_R
\, , \nn \\
\text{T$_3$-duality}: && \ \ \
\eps_R \ &\to \ \Gamma^3 \Gamma \eps_R
\, , \nn \\
&&
\pm \eps_R \ &= \ 
\Gamma \Gamma^3 \Gamma^{012345} \Gamma^3 \Gamma \eps_R
\ = \ 
- \Gamma^{012345} \eps_R
\, , \nn \\
\therefore \ \ \ 
\text{IIA $2^{(3,3)}_4(12,789,345)$} : && \ \ \ 
\pm \eps_L \ &= \ 
\Gamma^{012345} \eps_L
\, , \ls
\pm \eps_R \ = \ 
- \Gamma^{012345} \eps_R
\, , \nn \\
&& \text{namely,} \ \ \ 
\pm \eps \ &= \ 
\Gamma^{012345} \Gamma \eps
\, .
\end{alignat}

\item $\text{IIA $2^{(3,3)}_4(12,789,345)$} \xrightarrow{\text{T}_2} \text{IIB $1^{(4,3)}_4(1,789,2345)$}$:
\begin{alignat}{2}
\text{IIA $2^{(3,3)}_4(12,789,345)$} : && \ \ \ 
\pm \eps \ &= \ 
\Gamma^{012345} \Gamma \eps
\, , \nn \\
&& \text{i.e.,} \ \ 
\pm \eps_L \ &= \ 
\Gamma^{012345} \eps_L
\, , \ls
\pm \eps_R \ = \ 
- \Gamma^{012345} \eps_R
\, , \nn \\
\text{T$_2$-duality}: && \ \ \
\eps_R \ &\to \ \Gamma^2 \Gamma \eps_R
\, , \nn \\
&&
\pm \eps_R \ &= \ 
- \Gamma \Gamma^2 \Gamma^{012345} \Gamma^2 \Gamma \eps_R
\ = \ 
\Gamma^{012345} \eps_R
\, , \nn \\
\therefore \ \ \ 
\text{IIB $1^{(4,3)}_4(1,789,2345)$} : && \ \ \ 
\pm \eps_L \ &= \ 
\Gamma^{012345} \eps_L
\, , \ls
\pm \eps_R \ = \ 
\Gamma^{012345} \eps_R
\, , \nn \\
&& \text{namely,} \ \ \ 
\pm \eps \ &= \ 
\Gamma^{012345} \eps
\, .
\end{alignat}

\item $\text{IIB $1^{(4,3)}_4(1,789,2345)$} \xrightarrow{\text{T}_1} \text{IIA $0^{(5,3)}_4(,789,12345)$}$:
\begin{alignat}{2}
\text{IIB $1^{(4,3)}_4(1,789,2345)$} : && \ \ \ 
\pm \eps \ &= \ 
\Gamma^{012345} \eps
\, , \nn \\
&& \text{i.e.,} \ \ 
\pm \eps_L \ &= \ 
\Gamma^{012345} \eps_L
\, , \ls
\pm \eps_R \ = \ 
\Gamma^{012345} \eps_R
\, , \nn \\
\text{T$_1$-duality}: && \ \ \
\eps_R \ &\to \ \Gamma^1 \Gamma \eps_R
\, , \nn \\
&&
\pm \eps_R \ &= \ 
\Gamma \Gamma^1 \Gamma^{012345} \Gamma^1 \Gamma \eps_R
\ = \ 
- \Gamma^{012345} \eps_R
\, , \nn \\
\therefore \ \ \ 
\text{IIA $0^{(5,3)}_4(,789,12345)$} : && \ \ \ 
\pm \eps_L \ &= \ 
\Gamma^{012345} \eps_L
\, , \ls
\pm \eps_R \ = \ 
- \Gamma^{012345} \eps_R
\, , \nn \\
&& \text{namely,} \ \ \ 
\pm \eps \ &= \ 
\Gamma^{012345} \Gamma \eps
\, .
\end{alignat}

\end{itemize}

\subsubsection*{Descendants from the IIB $4^{(1,3)}_3$-brane}

\begin{itemize}
\item $\text{IIB $4^{(1,3)}_3(1234,567,9)$} \xrightarrow{\text{S}} \text{IIB $4^{(1,3)}_4 (1234,567,9)$}$: 
\begin{alignat}{2}
\text{IIB $4^{(1,3)}_3(1234,567,9)$}: && \ \ \ 
\pm \eps \ &= \ 
\Gamma^{012349} (\sigma_1) \eps
\, , \nn \\
\text{S-duality}: && \ \ \
\eps \ &\to \ S \eps
\, , \nn \\
\therefore \ \ \ 
\text{IIB $4^{(1,3)}_4(1234,567,9)$} : && \ \ \ 
\pm \eps \ &= \ 
\Gamma^{012349} S^{-1} (\sigma_1) S \eps
\nn \\
&&&= \ 
\Gamma^{012349} (\sigma_3) \eps
\, . 
\end{alignat}

\item $\text{IIB $4^{(1,3)}_4(1234,567,9)$} \xrightarrow{\text{T}_9} \text{IIA $5^3_4 (12349,567)$}$: 
\begin{alignat}{2}
\text{IIB $4^{(1,3)}_4(1234,567,9)$} : && \ \ \ 
\pm \eps \ &= \ 
\Gamma^{012349} (\sigma_3) \eps
\, , \nn \\
&& \text{i.e.,} \ \ 
\pm \eps_L \ &= \ 
\Gamma^{012349} \eps_L
\, , \ls
\pm \eps_R \ = \ 
- \Gamma^{012349} \eps_R
\, , \nn \\
\text{T$_9$-duality}: && \ \ \
\eps_R \ &\to \ \Gamma^9 \Gamma \eps_R
\, , \nn \\
&&
\pm \eps_R \ &= \ 
- \Gamma \Gamma^9 \Gamma^{012349} \Gamma^9 \Gamma \eps_R
\ = \ 
\Gamma^{012349} \eps_R
\, , \nn \\
\therefore \ \ \ 
\text{IIA $5^3_4(12349,567)$} : && \ \ \ 
\pm \eps_L \ &= \ 
\Gamma^{012349} \eps_L
\, , \ls
\pm \eps_R \ = \ 
\Gamma^{012349} \eps_R
\, , \nn \\
&& \text{namely,} \ \ \ 
\pm \eps \ &= \ 
\Gamma^{012349} \eps
\, .
\end{alignat}

\item $\text{IIB $4^{(1,3)}_4(1234,567,9)$} \xrightarrow{\text{T}_4} \text{IIA $3^{(2,3)}_4 (123,567,49)$}$: 
\begin{alignat}{2}
\text{IIB $4^{(1,3)}_4(1234,567,9)$} : && \ \ \ 
\pm \eps \ &= \ 
\Gamma^{012349} (\sigma_3) \eps
\, , \nn \\
&& \text{i.e.,} \ \ 
\pm \eps_L \ &= \ 
\Gamma^{012349} \eps_L
\, , \ls
\pm \eps_R \ = \ 
- \Gamma^{012349} \eps_R
\, , \nn \\
\text{T$_4$-duality}: && \ \ \
\eps_R \ &\to \ \Gamma^4 \Gamma \eps_R
\, , \nn \\
&&
\pm \eps_R \ &= \ 
- \Gamma \Gamma^4 \Gamma^{012349} \Gamma^4 \Gamma \eps_R
\ = \ 
\Gamma^{012349} \eps_R
\, , \nn \\
\therefore \ \ \ 
\text{IIA $3^{(2,3)}_4(123,567,49)$} : && \ \ \ 
\pm \eps_L \ &= \ 
\Gamma^{012349} \eps_L
\, , \ls
\pm \eps_R \ = \ 
\Gamma^{012349} \eps_R
\, , \nn \\
&& \text{namely,} \ \ \ 
\pm \eps \ &= \ 
\Gamma^{012349} \eps
\, .
\end{alignat}

\item $\text{IIA $3^{(2,3)}_4(123,567,49)$} \xrightarrow{\text{T}_3} \text{IIB $2^{(3,3)}_4 (12,567,349)$}$: 
\begin{alignat}{2}
\text{IIA $3^{(2,3)}_4(123,567,49)$} : && \ \ \ 
\pm \eps \ &= \ 
\Gamma^{012349} \eps
\, , \nn \\
&& \text{i.e.,} \ \ 
\pm \eps_L \ &= \ 
\Gamma^{012349} \eps_L
\, , \ls
\pm \eps_R \ = \ 
\Gamma^{012349} \eps_R
\, , \nn \\
\text{T$_3$-duality}: && \ \ \
\eps_R \ &\to \ \Gamma^3 \Gamma \eps_R
\, , \nn \\
&&
\pm \eps_R \ &= \ 
\Gamma \Gamma^3 \Gamma^{012349} \Gamma^3 \Gamma \eps_R
\ = \ 
- \Gamma^{012349} \eps_R
\, , \nn \\
\therefore \ \ \ 
\text{IIB $2^{(3,3)}_4(12,567,349)$} : && \ \ \ 
\pm \eps_L \ &= \ 
\Gamma^{012349} \eps_L
\, , \ls
\pm \eps_R \ = \ 
- \Gamma^{012349} \eps_R
\, , \nn \\
&& \text{namely,} \ \ \ 
\pm \eps \ &= \ 
\Gamma^{012349} (\sigma_3) \eps
\, .
\end{alignat}

\item $\text{IIB $2^{(3,3)}_4(12,567,349)$} \xrightarrow{\text{T}_2} \text{IIA $1^{(4,3)}_4 (1,567,2349)$}$: 
\begin{alignat}{2}
\text{IIB $2^{(3,3)}_4(12,567,349)$} : && \ \ \ 
\pm \eps \ &= \ 
\Gamma^{012349} (\sigma_3) \eps
\, , \nn \\
&& \text{i.e.,} \ \ 
\pm \eps_L \ &= \ 
\Gamma^{012349} \eps_L
\, , \ls
\pm \eps_R \ = \ 
- \Gamma^{012349} \eps_R
\, , \nn \\
\text{T$_2$-duality}: && \ \ \
\eps_R \ &\to \ \Gamma^2 \Gamma \eps_R
\, , \nn \\
&&
\pm \eps_R \ &= \ 
- \Gamma \Gamma^2 \Gamma^{012349} \Gamma^2 \Gamma \eps_R
\ = \ 
\Gamma^{012349} \eps_R
\, , \nn \\
\therefore \ \ \ 
\text{IIA $1^{(4,3)}_4(1,567,2349)$} : && \ \ \ 
\pm \eps_L \ &= \ 
\Gamma^{012349} \eps_L
\, , \ls
\pm \eps_R \ = \ 
\Gamma^{012349} \eps_R
\, , \nn \\
&& \text{namely,} \ \ \ 
\pm \eps \ &= \ 
\Gamma^{012349} \eps
\, .
\end{alignat}

\item $\text{IIA $1^{(4,3)}_4(1,567,2349)$} \xrightarrow{\text{T}_1} \text{IIB $0^{(5,3)}_4 (,567,12349)$}$: 
\begin{alignat}{2}
\text{IIA $1^{(4,3)}_4(1,567,2349)$} : && \ \ \ 
\pm \eps \ &= \ 
\Gamma^{012349} \eps
\, , \nn \\
&& \text{i.e.,} \ \ 
\pm \eps_L \ &= \ 
\Gamma^{012349} \eps_L
\, , \ls
\pm \eps_R \ = \ 
\Gamma^{012349} \eps_R
\, , \nn \\
\text{T$_1$-duality}: && \ \ \
\eps_R \ &\to \ \Gamma^1 \Gamma \eps_R
\, , \nn \\
&&
\pm \eps_R \ &= \ 
\Gamma \Gamma^1 \Gamma^{012349} \Gamma^1 \Gamma \eps_R
\ = \ 
- \Gamma^{012349} \eps_R
\, , \nn \\
\therefore \ \ \ 
\text{IIB $0^{(5,3)}_4(,567,23349)$} : && \ \ \ 
\pm \eps_L \ &= \ 
\Gamma^{012349} \eps_L
\, , \ls
\pm \eps_R \ = \ 
- \Gamma^{012349} \eps_R
\, , \nn \\
&& \text{namely,} \ \ \ 
\pm \eps \ &= \ 
\Gamma^{012349} (\sigma_3) \eps
\, .
\end{alignat}

\end{itemize}

\subsection{Uplifting to M-theory}
\label{app:IIA2M}

Here we uplift type IIA branes to those of M-theory.
The mass formulae of type IIA branes are rewritten as those of M-theory branes via the relation (\ref{mass-IIA2M}).

\subsubsection*{Uplifting IIA solitonic five-branes}

\begin{itemize}
\item IIA NS5(12345) $\xrightarrow{\text{uplift}}$ M5(12345):
\begin{alignat}{2}
M_{\text{NS5}} \ &= \ 
\frac{R_1 R_2 R_3 R_4 R_5}{g_s^2 \ell_s^6}
\, , &\ls
\pm \eps \ &= \ \Gamma^{012345} \eps
\, , \nn \\
\to \ \ \
M_{\text{M5}} \ &= \ \frac{R_1 R_2 R_3 R_4 R_5}{\ell_P^6}
\, , &\ls
\pm \eta \ &= \ \Gamma^{012345} \eta
\, . 
\end{alignat}

\item IIA KK5(12345,9) $\xrightarrow{\text{uplift}}$ KK6(12345$\natural$,9):
\begin{alignat}{2}
M_{\text{KK5}} \ &= \ 
\frac{R_1 R_2 R_3 R_4 R_5 (R_9)^2}{g_s^2 \ell_s^8}
\, , &\ls
\pm \eps \ &= \ \Gamma^{012345} \Gamma \eps
\, , \nn \\
\to \ \ \
M_{\text{KK6}} \ &= \ 
\frac{R_1 R_2 R_3 R_4 R_5 R_{\natural} (R_9)^2}{\ell_P^9}
\, , &\ls
\pm \eta \ &= \ \Gamma^{012345\natural} \eta
\, . 
\end{alignat}

\item IIA $5^2_2(12345,89)$ $\xrightarrow{\text{uplift}}$ $5^3(12345,89\natural)$:
\begin{alignat}{2}
M_{5^2_2} \ &= \ 
\frac{R_1 R_2 R_3 R_4 R_5 (R_8 R_9)^2}{g_s^2 \ell_s^{10}}
\, , &\ls
\pm \eps \ &= \ \Gamma^{012345} \eps
\, , \nn \\
\to \ \ \
M_{5^3} \ &= \ 
\frac{R_1 R_2 R_3 R_4 R_5 (R_8 R_9 R_{\natural})^2}{\ell_P^{12}}
\, , &\ls
\pm \eta \ &= \ \Gamma^{012345} \eta
\, . 
\end{alignat}

\item IIA $5^3_2(12345,789)$ $\xrightarrow{\text{uplift}}$ $5^{(1,3)}(12345,789,\natural)$:
\begin{alignat}{2}
M_{5^3_2} \ &= \ 
\frac{R_1 R_2 R_3 R_4 R_5 (R_7 R_8 R_9)^2}{g_s^2 \ell_s^{12}}
\, , &\ls
\pm \eps \ &= \ \Gamma^{012345} \Gamma \eps
\, , \nn \\
\to \ \ \
M_{5^{(1,3)}} \ &= \ 
\frac{R_1 R_2 R_3 R_4 R_5 (R_7 R_8 R_9)^2 (R_{\natural})^3}{\ell_P^{15}}
\, , &\ls
\pm \eta \ &= \ \Gamma^{012345\natural} \eta
\, . 
\end{alignat}

\end{itemize}

\subsubsection*{Uplifting IIA defect branes}

We also uplift the defect branes in type IIA theory (\ref{SPR-IIA-defect}).
\begin{itemize}
\item $1^6_4(1,234567)$ $\xrightarrow{\text{uplift}}$ $2^{6}(1\natural,234567)$:
\begin{alignat}{2}
M_{1^6_3} \ &= \ 
\frac{R_1 (R_2 R_3 R_4 R_5 R_6 R_7)^2}{g_s^4 \ell_s^{14}}
\, , &\ls
\pm \eps \ &= \ \Gamma^{01} \Gamma \eps
\, , \nn \\
\to \ \ \
M_{2^6} \ &= \ 
\frac{R_1 R_{\natural} (R_2 R_3 R_4 R_5 R_6 R_7)^2}{\ell_P^{15}}
\, , &\ls
\pm \eta \ &= \ \Gamma^{01\natural} \eta
\, . 
\end{alignat}

\item $0^{(1,6)}_4(,234567,1)$ $\xrightarrow{\text{uplift}}$ $0^{(1,7)}(,234567\natural,1)$:
\begin{alignat}{2}
M_{0^{(1,6)}_4} \ &= \ 
\frac{(R_2 R_3 R_4 R_5 R_6 R_7)^2 (R_1)^3}{g_s^4 \ell_s^{16}}
\, , &\ls
\pm \eps \ &= \ \Gamma^{01} \eps
\, , \nn \\
\to \ \ \
M_{0^{(1,7)}} \ &= \ 
\frac{(R_2 R_3 R_4 R_5 R_6 R_7 R_{\natural})^2 (R_1)^3}{\ell_P^{18}}
\, , &\ls
\pm \eta \ &= \ \Gamma^{01} \eta
\, . 
\end{alignat}

\end{itemize}
Other defect branes in (\ref{SPR-IIA-defect}) are uplifted to some of the (exotic) branes in M-theory that have already appeared.

\end{appendix}

}

\begin{thebibliography}{99}

\bibitem{Blau:1997du}
  M.~Blau and M.~O'Loughlin,
  ``{\sl Aspects of U-duality in matrix theory},''
  Nucl.\ Phys.\ B {\bf 525} (1998) 182
  [hep-th/9712047].

\bibitem{Obers:1998fb}
  N.~A.~Obers and B.~Pioline,
  ``{\sl U-duality and M-theory},''
  Phys.\ Rept.\  {\bf 318} (1999) 113
  [hep-th/9809039].

\bibitem{Eyras:1999at}
  E.~Eyras and Y.~Lozano,
  ``{\sl Exotic branes and nonperturbative seven-branes},''
  Nucl.\ Phys.\ B {\bf 573} (2000) 735
  [hep-th/9908094].

\bibitem{LozanoTellechea:2000mc}
  E.~Lozano-Tellechea and T.~Ort\'{\i}n,
  ``{\sl 7-branes and higher Kaluza-Klein branes},''
  Nucl.\ Phys.\ B {\bf 607} (2001) 213
  [hep-th/0012051].

\bibitem{deBoer:2010ud}
  J.~de Boer and M.~Shigemori,
  ``{\sl Exotic branes and non-geometric backgrounds},''
  Phys.\ Rev.\ Lett.\  {\bf 104} (2010) 251603
  [arXiv:1004.2521 [hep-th]].


\bibitem{Giveon:1998sr}
  A.~Giveon and D.~Kutasov,
  ``{\sl Brane dynamics and gauge theory},''
  Rev.\ Mod.\ Phys.\  {\bf 71} (1999) 983
  [hep-th/9802067].

\bibitem{Greene:1989ya}
  B.~R.~Greene, A.~D.~Shapere, C.~Vafa and S.~T.~Yau,
  ``{\sl Stringy cosmic strings and noncompact Calabi-Yau manifolds},''
  Nucl.\ Phys.\ B {\bf 337} (1990) 1.

\bibitem{Bergshoeff:2006jj}
  E.~A.~Bergshoeff, J.~Hartong, T.~Ort\'{\i}n and D.~Roest,
  ``{\sl Seven-branes and Supersymmetry},''
  JHEP {\bf 0702} (2007) 003
  [hep-th/0612072].

\bibitem{Kikuchi:2012za}
  T.~Kikuchi, T.~Okada and Y.~Sakatani,
  ``{\sl Rotating string in doubled geometry with generalized isometries},''
  Phys.\ Rev.\ D {\bf 86} (2012) 046001
  [arXiv:1205.5549 [hep-th]].

\bibitem{deBoer:2012ma}
  J.~de Boer and M.~Shigemori,
  ``{\sl Exotic branes in string theory},''
  Phys.\ Rept.\  {\bf 532} (2013) 65
  [arXiv:1209.6056 [hep-th]].

\bibitem{Kimura:2014wga}
  T.~Kimura,
  ``{\sl Defect $(p,q)$ five-branes},''
  Nucl.\ Phys.\ B {\bf 893} (2015) 1
  [arXiv:1410.8403 [hep-th]].

\bibitem{Kimura:2014bea}
  T.~Kimura, S.~Sasaki and M.~Yata,
  ``{\sl Hyper-K\"{a}hler with torsion, T-duality, and defect $(p,q)$ five-branes},''
  JHEP {\bf 1503} (2015) 076
  [arXiv:1411.3457 [hep-th]].

\bibitem{Okada:2014wma}
  T.~Okada and Y.~Sakatani,
  ``{\sl Defect branes as Alice strings},''
  JHEP {\bf 1503} (2015) 131
  [arXiv:1411.1043 [hep-th]].

\bibitem{Bergshoeff:2011se}
  E.~A.~Bergshoeff, T.~Ort\'{\i}n and F.~Riccioni,
  ``{\sl Defect branes},''
  Nucl.\ Phys.\ B {\bf 856} (2012) 210
  [arXiv:1109.4484 [hep-th]].

\bibitem{Sakatani:2014hba}
  Y.~Sakatani,
  ``{\sl Exotic branes and non-geometric fluxes},''
  JHEP {\bf 1503} (2015) 135
  [arXiv:1412.8769 [hep-th]].


\bibitem{Kleinschmidt:2011vu}
  A.~Kleinschmidt,
  ``{\sl Counting supersymmetric branes},''
  JHEP {\bf 1110} (2011) 144
  [arXiv:1109.2025 [hep-th]].

\bibitem{Andriot:2013xca}
  D.~Andriot and A.~Betz,
  ``{\sl $\beta$-supergravity: a ten-dimensional theory with non-geometric fluxes, and its geometric framework},''
  JHEP {\bf 1312} (2013) 083
  [arXiv:1306.4381 [hep-th]].

\bibitem{Andriot:2014uda}
  D.~Andriot and A.~Betz,
  ``{\sl NS-branes, source corrected Bianchi identities, and more on backgrounds with non-geometric fluxes},''
  JHEP {\bf 1407} (2014) 059
  [arXiv:1402.5972 [hep-th]].

\bibitem{Blair:2014zba}
  C.~D.~A.~Blair and E.~Malek,
  ``{\sl Geometry and fluxes of $SL(5)$ exceptional field theory},''
  JHEP {\bf 1503} (2015) 144
  [arXiv:1412.0635 [hep-th]].

\bibitem{Gauntlett:1997cv}
  J.~P.~Gauntlett,
  ``{\sl Intersecting branes},''
  In Seoul/Sokcho 1997, Dualities in gauge and string theories, pp.146-193
  [hep-th/9705011].

\bibitem{Schwarz:1983qr}
  J.~H.~Schwarz,
  ``{\sl Covariant field equations of chiral $N=2$ $D=10$ supergravity},''
  Nucl.\ Phys.\ B {\bf 226} (1983) 269.

\bibitem{Imamura}
Y.~Imamura,
``{\sl String, M and Matrix theories},''
Soryushiron Kenkyu {\bf 96}-5 (1998) 187
(in Japanese).


\bibitem{Bergshoeff:2012pm}
  E.~A.~Bergshoeff, A.~Kleinschmidt and F.~Riccioni,
  ``{\sl Supersymmetric domain walls},''
  Phys.\ Rev.\ D {\bf 86} (2012) 085043
  [arXiv:1206.5697 [hep-th]].

\end{thebibliography}
\end{document}